\documentclass[ba]{imsart}

\RequirePackage{amsthm,amsmath,amsfonts,amssymb}
\RequirePackage[numbers]{natbib}
\RequirePackage[colorlinks,citecolor=blue,urlcolor=blue,backref=page,backref=page]{hyperref}
\RequirePackage{graphicx}
\usepackage{multirow}
\usepackage{booktabs}

\pubyear{2024}
\volume{TBA}
\issue{TBA}
\firstpage{1}
\lastpage{1}

\startlocaldefs
\theoremstyle{plain}

\newtheorem{theorem}{Theorem}[section]

\newtheorem{remark}{Remark}
\theoremstyle{definition}

\theoremstyle{remark}

\DeclareMathOperator*{\argmax}{arg\,max}
\newcommand{\bea}{\begin{eqnarray*}}
	\newcommand{\eea}{\end{eqnarray*}}
\newcommand{\bean}{\begin{eqnarray}}
	\newcommand{\eean}{\end{eqnarray}}
\newcommand{\lra}{\longrightarrow}

\newcommand{\sg}{\Sigma}

\newcommand{\calD}{\mathcal{D}}

\newcommand{\calL}{\mathcal{L}}

\newcommand{\bbR}{\mathbb{R}}
\newcommand{\bbE}{\mathbb{E}}

\allowdisplaybreaks
\endlocaldefs

\begin{document}

\begin{frontmatter}
\title{Scalable Bayesian inference on high-dimensional multivariate linear regression}
\runtitle{Bayesian multivariate linear regression}

\begin{aug}
\author[A]{\fnms{Xuan}~\snm{Cao}\ead[label=e1]{xuan.cao@uc.edu}}
\and
\author[B]{\fnms{Kyoungjae}~\snm{Lee}\ead[label=e2]{leekjstat@gmail.com}}

\address[A]{Department of Mathematical Sciences,
	University of Cincinnati, Cincinnati, Ohio, USA\printead[presep={,\ }]{e1}}

\address[B]{Corresponding author. Department of Statistics, 
	Sungkyunkwan University, Seoul, South Korea\printead[presep={,\ }]{e2}}
\runauthor{X. Cao and K. Lee}
\end{aug}

\begin{abstract}
We consider jointly estimating the coefficient matrix and the error precision matrix in high-dimensional multivariate linear regression models. 
Bayesian methods in this context often face computational challenges, leading to previous approaches that either utilize a generalized likelihood without ensuring the positive definiteness of the precision matrix or rely on maximization algorithms targeting only the posterior mode, thus failing to address uncertainty.
In this work, we propose two Bayesian methods: an exact method and an approximate two-step method.
We first propose an exact method based on spike and slab priors for the coefficient matrix and DAG-Wishart prior for the error precision matrix, whose computational complexity is comparable to the state-of-the-art generalized likelihood-based Bayesian method. 
To further enhance scalability, a two-step approach is developed by ignoring the dependency structure among response variables.
This method estimates the coefficient matrix first, followed by the calculation of the posterior of the error precision matrix based on the estimated errors.
We validate the two-step method by demonstrating (i) selection consistency and posterior convergence rates for the coefficient matrix and (ii) selection consistency for the directed acyclic graph (DAG) of errors.
We demonstrate the practical performance of proposed methods through synthetic and real data analysis.
\end{abstract}

\begin{keyword}[class=MSC]
\kwd[Primary ]{62F15}
\kwd{62H12}
\kwd[; secondary ]{62F12}
\end{keyword}

\begin{keyword}
\kwd{Generalized likelihood}
\kwd{Selection consistency}
\kwd{Posterior convergence rate}
\end{keyword}

\end{frontmatter}

\section{Introduction}\label{sec:intro}


Multivariate linear regression models have attracted attention in the statistical literature over the past two decades because they provide an intuitive tool for capturing the relationships between a response vector and a covariate vector. 
Unlike in univariate linear regression models, learning the dependency structure among response variables becomes a critical task in multivariate linear regression models. 
In particular, when there is time ordering or known causal relationships among the response variables, it is crucial to incorporate this information into the inference. For example, consider time series data of a climate variable observed at specific locations as a response vector, with the location information and other explanatory variables as a covariate vector. 
Using data from multiple locations, a multivariate linear regression model can be applied to infer how location and covariates influence specific climate variables. 
In this case, since there is time ordering among the response variables, it is natural to reflect this in the inference.
In this paper, we propose a Bayesian method for multivariate linear regression models that can incorporate such known ordering among the response variables. 
We are particularly interested in methods that allow for joint inference of the coefficient matrix, say $B$, and the precision matrix of the error term, say $\Omega$, especially in high-dimensional settings.

Several frequentist approaches to solve the joint estimation problem have been proposed and studied in the literature, 
see \cite{RLZ:2010b, Sohn:Kim:2012, Yin:Li:2013, CLLX:2013, Mccarter:Kim:2014, Yuan:Zhang:2014, MRZZ:2016, LBBM:2016} and the references therein. 
To address the high-dimensional setting, 
these methods induce sparsity in the regression coefficient matrix and the precision matrix. 
This is achieved by optimizing over objective functions comprising of a data-based loss function and an appropriate penalty function to induce sparsity. 
For some of these methods, for example \cite{MRZZ:2016} and \cite{LBBM:2016}, algorithmic and statistical convergence properties have also been thoroughly studied.

On the Bayesian side, there have been mostly two schools of work, those based on the exact likelihood and others based on the generalized likelihood. 
Most of the exact likelihood-based methods, except for \cite{Deshpande2019}, have at least one of these two limitations: (i) assuming $B$ to be row-wise sparse \citep{BM:2013, CLP:2017, Dai2022} and/or (ii) restricting their analysis to only a special class of precision matrices/graphs such as decomposable graphs \citep{BM:2013, Bottolo2021}. 
Row-wise sparsity of $B$ means that if a certain covariate is not active, then the entire corresponding row in the coefficient matrix is zero.
This implies that each covariate either affects all response variables or affects none of them.
In contrast to the above approaches, \cite{Deshpande2019} assumed entry-wise sparsity of $B$ and a general sparse positive definite matrix $\Omega$ by employing multivariate spike and slab Lasso priors. 
However, their method only provided point estimates without generating posterior samples, making uncertainty quantification through credible intervals impossible.
\cite{Ha2021} and \cite{Samanta2022} subsequently proposed the methods for entry-wise selection in $B$ and $\Omega$ based on generalized likelihoods for scalable posterior inference, but none of their methods ensure the positive definiteness of $\Omega$.


Motivated by the gap in the literature, we develop two computationally scalable Bayesian methods for the selection and estimation of sparse $B$ and $\Omega$, while incorporating the ordering among the response variables. 
We use the modified Cholesky decomposition (MCD) for $\Omega$, which allows us to leverage the ordering information, achieve fast computation and ensure the positive definiteness of $\Omega$.
The first approach relies on the exact likelihood.
We adopt discrete spike and slab priors and DAG-Wishart prior to capture significant signals in $B$ and Cholesky factor of $\Omega$, respectively.
Its computational cost per Markov chain Monte Carlo (MCMC) iteration is comparable to that of the fastest among existing Bayesian methods \citep{Samanta2022}.
However, it can still be computationally expensive in high-dimensional settings. 
To further enhance computational speed, inspired by \cite{Samanta2022} and \cite{CLLX:2013}, we propose a two-step approach. 
In the first step, we simplify the multivariate linear regression model into $q$ independent univariate linear regression models by ignoring the dependencies among the response variables, and apply a variant of spike and slab prior to each coefficient vector to infer $B$. 
Using the obtained estimate of $B$, we calculate the estimated errors by removing the effects of covariates from the responses. 
In the second step, we assume these estimated errors as observations in a Gaussian model and apply the DAG-Wishart prior to infer the Cholesky factor of $\Omega$.
The computational cost of the proposed two-step method is lower than that of any existing Bayesian methods, making it scalable even in high-dimensional settings.
We establish model selection consistency and posterior contraction rates for the two-step approach under mild regularity conditions (Theorems \ref{thm:selection_gamma}--\ref{thm:strong_sel_L}), and simulation studies further signify its satisfactory empirical performance. 
The associated computational gains, along with the theoretical properties, render the two-step approach suitable and preferable in high dimensions.

While leveraging the MCD of $\Omega$ ensures its positive definiteness, we assume an ordering among the response variables.
Therefore, the developed method can be naturally applied in cases where the data has an ordering, such as time ordering or causal relationships.
In the absence of ordering information, methods to estimate the ordering should be applied before using the approach discussed in this paper. Further details can be found in \cite{harris2013pc} and \cite{park2021learning}.

The rest of paper is organized as follows. 
Section \ref{sec2} introduces the multivariate linear regression model, as well as the concepts of the Gaussian directed acyclic graph (DAG) model and the DAG-Wishart distribution, which are necessary to formulate the proposed model in this paper.
In Section~\ref{sec:model}, we present a Bayesian method based on the exact likelihood and its posterior computation.
In Section~\ref{sec:two-step}, we introduce a scalable two-step Bayesian method based on generalized likelihoods, along with its theoretical guarantees.
The practical performance of the proposed Bayesian methods is demonstrated through simulation studies and real data analysis in Sections \ref{sec:simul} and \ref{sec:real}, respectively. 
Finally, in Section~\ref{sec:disc}, we provide a guideline for selecting between the two proposed Bayesian methods and discuss several potential directions for future research.

\section{Preliminaries}\label{sec2}

\subsection{Multivariate linear regression model}\label{subsec:mtv_linear}

	We consider the following multivariate linear regression models, 
	\bean\label{model:Yi}
	\begin{split}
		Y_i  \,\,&=\,\, B^T X_i + E_i , \quad i=1,\ldots, n  ,
	\end{split}
	\eean
	where $Y_i =(y_{i1}, \ldots, y_{iq})^T\in \bbR^q$ is the $i$th observation, $X_i =(x_{i1},\ldots, x_{ip} )^T \in\bbR^p$ is the $i$th covariate vector, and
	$B = (b_1, \ldots, b_q) = (B_1, \ldots, B_p)^T \in \bbR^{p\times q}$ is the coefficient matrix. 
	Let $Y = (Y_1,\ldots, Y_n)^T = (y_1,\ldots, y_q)\in\bbR^{n\times q}$ be the data matrix, where $y_j = \left(y_{1j}, \ldots, y_{nj}\right)^T$ is the observed value on the $j$th variable.
	Let $X= (x_1, \ldots, x_p) = (X_1,\ldots, X_n)^T \in\bbR^{n\times p}$ be the design matrix.
	
	For the $i$th error vector $E_i$, we assume $E_i \mid \Sigma \,\,\overset{iid}{\sim}\,\, N_q(0, \Sigma)$, where $\sg\in\bbR^{q\times q}$ is the covariance matrix.
	We assume that the error $E_i$ in \eqref{model:Yi} follows the Gaussian DAG model, whose details will be described in the following subsection.

	\subsection{Gaussian DAG model and DAG-Wishart prior} \label{sec2.1}
	
	We denote a DAG as $\mathcal{D} = (V,E_{\rm ed})$, where $V = \{1,\ldots,q\}$  and  $E_{\rm ed}$ are a vertex set and an edge set, respectively. 
	In the edge set, there is no directed path that originates and terminates at the same vertex.
	As in \cite{BLMR:2016}, we assume that a parent ordering is known and  all the edges are directed from larger vertices to smaller vertices. 
	For $j = 1, \ldots, q$, the set of parents of $j$, denoted by $pa_j(\mathcal D)$, is the collection of all vertices which are larger than $j$ and share an edge with $j$. 
	We denote a Gaussian DAG model over a given DAG $\mathcal{D}$ as  $\mathcal{N}_{\mathcal{D}}$ and define it as the set of all multivariate Gaussian distributions which obey the directed Markov property with respect to a DAG $\mathcal{D}$. 
	In particular, 
	if ${Z}=(Z_1, \ldots, Z_q)^T \sim N_q\big(0,\Sigma \big)$ and $N_q(0,\Sigma) \in 
	\mathcal{N}_{\mathcal{D}}$, then $Z_j \perp   {Z}_{\{j+1,\ldots,q\}\backslash pa_j(\mathcal D)}|
	{Z}_{pa_j(\mathcal D)}$ for each $j = 1, \ldots, q-1$. 
	
	By the MCD, any positive definite matrix $\Omega\in\bbR^{q\times q}$ can be uniquely decomposed as $\Omega = LD^{-1}L^T$, where $L\in\bbR^{q\times q}$ is a lower triangular matrix with unit diagonal entries, and $D = \mbox{Diag}(d_1, \ldots, d_q) \in\bbR^{q\times q}$ is a diagonal matrix with $d_j>0$ \citep{Pourahmadi:2007}.
	For a given DAG $\mathcal{D}$ on $q$ vertices, we denote $\mathcal{L}_{\mathcal{D}}$ as the set of lower triangular matrices with unit diagonals and $L_{ij} = 0$ whenever $i \notin pa_j(\mathcal D)$ for $1\le j < i \le q$. 
	Let $\mathcal{D}_+^q$ be the set of $q\times q$ diagonal matrices with  positive diagonal entries. 
	We refer to $\Theta_{\mathcal{D}} = \mathcal{L}_{\mathcal{D}}\times \mathcal{D}_+^q$ as the Cholesky space corresponding to $\mathcal{D}$. 
	For a given DAG $\calD$, it is well-known that $\mathcal{N}_{\mathcal{D}} = \big\{N_q(0,  (LD^{-1} L^T)^{-1} ):(L,D) \in 
	\Theta_{\mathcal{D}}\big\}$.
	Therefore, by considering a multivariate Gaussian distribution with a sparse Cholesky factor $L \in \calL_{\calD}$, it is equivalent to adopt a Gaussian DAG model with a corresponding DAG $\calD$.

	Note that, in model \eqref{model:Yi}, we assume $E_i \mid \sg \overset{iid}{\sim} N_q(0,\sg)$.
	Let $\Omega=(\omega_{kj}) =\sg^{-1}= L    D^{-1}   L^T$ be the MCD of  $\Omega $, then it is equivalent to the following Gaussian DAG model
	\bean\label{model:Ei}
	E_i\mid L,D ,\calD  &\overset{iid}{\sim}& N_q(0, (LD^{-1} L^T)^{-1} ),\quad  i=1,\ldots, n,
	\eean 
	where $(L,D) \in \Theta_{\calD}$ for an unknown DAG $\calD$.	 
	Note that \eqref{model:Ei} can be represented as a sequence of linear regression models, which facilitates posterior inference. 
	For more details, refer to the supplementary material or \cite{lee2019minimax}.


		
		Now, we introduce the hierarchical DAG-Wishart prior  \citep{BLMR:2016} for $\mathcal{D}$ and $(L,D) \in \Theta_{\mathcal{D}}$.
		Let $\mathcal{D}$ be a DAG on $q$ vertices and $A\in\bbR^{q\times q}$ be a symmetric matrix.
		For a given $j  = 1, \ldots, q-1$, we define a submatrix of $A$ as a block matrix, 
		$$ A_{\mathcal D}^{ \ge j} = 
		\left[ \begin{matrix}
			A_{jj} & (A_{\mathcal D.j}^>)^T \\
			A_{\mathcal D.j}^> & A_{\mathcal D}^{>j}
		\end{matrix} \right]  \in \mathbb{R}^{(\nu_{j}(\mathcal{D})+1) \times (\nu_{j}(\mathcal{D}) +1) },
		$$
		where  $\nu_j(\calD) = |pa_j(\calD)|$, $A_{\mathcal{D} .j}^> = (A_{rj})_{r \in pa_{j}(\mathcal{D})} \in \mathbb{R}^{\nu_{j}(\mathcal{D})}$ is a subvector consisting of the rows corresponding to $pa_j(\mathcal{D})$ in the $j$th column of $A$, and
		$A_{\mathcal{D}}^{>j} = (A_{rs})_{r,s \in pa_{j}(\mathcal{D})} \in \mathbb{R}^{\nu_{j}(\mathcal{D}) \times \nu_{j}(\mathcal{D})}$ is a submatrix of $A$ consisting of the rows and columns  corresponding to $pa_j(\mathcal{D})$. 
		Let $A_{\mathcal D.j}^{\ge} = (A_{jj}, (A_{\mathcal D.j}^>)^T)^T\in \bbR^{\nu_{j}(\mathcal D)+1}$ be the first column vector of $A_{\mathcal{D}}^{\ge j}$. 
		For notational convenience, let $A_{\mathcal D.q}^{\ge} = A_{\mathcal D}^{ \ge q} = A_{qq}$.

		For $\mathcal{D}$ and $(L,D) \in \Theta_{\mathcal{D}}$, we assume the following hierarchical DAG-Wishart prior,
		\bean
		\pi (\mathcal D) &=& \prod_{j=1}^{q-1} \eta_2^{\nu_j(\mathcal{D})} (1-\eta_2)^{q-j-
			\nu_j(\mathcal{D})}, \label{model:dag}  \\
		(   L,    D) \mid  \mathcal D &\sim& \frac{1}{z_{\mathcal{D}}(U,{\phi})} \exp\Big\{-\frac12\mbox{tr}
		\big((LD^{-1}L^T)U\big)\Big\} \prod_{j=1}^q d_{j}^{-\frac{\phi_j(\mathcal{D})}2}  ,  \label{model:LD}
		\eean
		for some  constant $0< \eta_2<1$, a positive definite matrix $U\in\bbR^{q\times q}$, and a $p$-dimensional vector ${\phi}(\calD)= ({\phi}_1(\calD),\ldots, {\phi}_q(\calD))^T\in\bbR^q$ with $\phi_j(\mathcal D) - \nu_j(\mathcal D) >2$ for all $j = 1, \ldots, q$.  
		For the prior density \eqref{model:dag} defined on the whole space of DAGs, we assume that each directed edge is present with probability $\eta_2$. 
		The normalizing constant for \eqref{model:LD} can be calculated as 
		\begin{align*}
			z_{\mathcal{D}}(U,{\phi}) \,= \,\prod_{j=1}^{q} \zeta_{\calD}^j(U, \phi) 
			\,=\,\prod_{j=1}^{q}
			\frac{\Gamma(\frac{\phi_j(\mathcal{D})}2 - \frac{\nu_{j}(\mathcal D)}2 - 
				1)2^{\frac{\phi_j(\mathcal{D})}2 - 1}(\sqrt{\pi})^{\nu_{j}(\mathcal D)} 
			}{\big(U_{j|pa_j({\mathcal{D}})}\big)^{\frac{\phi_j(\mathcal{D})}2 - \frac{\nu_{j}
						(\mathcal D)}2 - 1} \det(U_{\mathcal D}^{>j})^{\frac 1 2} } ,
		\end{align*}
		where  $U_{j|pa_j({\mathcal{D}})} = U_{jj} - (U_{\mathcal D \cdot j}^>)^T 
		(U_{\mathcal{D}}^{>j})^{-1} U_{\mathcal D \cdot j}^>$ stands for the Schur complement of the block $U_{\mathcal D}^{>j}$ of the matrix $U_{\mathcal D}^{\ge j}$ \citep{BLMR:2016}.

		\begin{remark}
			For a given undirected graph $G$, which determines the support of the precision matrix $\Omega$, the $G$-Wishart distribution \citep{roverato2002hyper,mohammadi2015bayesian} has been used as a prior for sparse precision matrices. 		
			Although the DAG-Wishart density \eqref{model:LD} is similar to the density of Wishart distributions, in general, it is known that the support of the Cholesky factor $L$ (edges in a DAG $\mathcal{D}$) does not exactly match the support of the precision matrix $\Omega$ (edges in an undirected graph $G$).		
			Furthermore, the DAG-Wishart distribution has multiple shape parameters ${\phi}(\mathcal{D}) \in\bbR^q$, which distinguishes it from the $G$-Wishart distribution.
		\end{remark}
		
		\begin{remark}
			The hyper Markov property of \eqref{model:LD} renders the mutual independence of $(L_{\mathcal D.j}^> , d_{j})$ \citep{BLMR:2016}.
			Specifically, if $(L,D)$ follows density \eqref{model:LD}, it is equivalent to say that $(L,D)$ follows the following conditional distributions, which are useful for computing conditional posteriors:
			\begin{eqnarray*}
				\begin{split}
					L_{\mathcal D.j}^> \mid d_{j}, pa_j(\mathcal D) &\overset{ind}{\sim} N_{\nu_j(\mathcal D)}\big(-(U_{\mathcal D}^{>j})^{-1}U_{\mathcal D.j}^{>},  \, d_{j}(U_{\mathcal D}^{>j})^{-1}\big),  \quad j =  1,\ldots, q-1,\\
					d_{j} \mid pa_j(\mathcal D) &\overset{ind}{\sim} \mbox{Inverse-Gamma}\big(\frac{\phi_j(\mathcal{D})-\nu_{i}(\mathcal D)}2  - 
					1, \frac 1 2 U_{j|pa_j({\mathcal{D}})}\big), \,\,  j =1,\ldots, q.
				\end{split}			
			\end{eqnarray*}
		\end{remark}

		\section{Exact likelihood-based Bayesian approach}\label{sec:model}

		By combining  \eqref{model:Yi} and \eqref{model:Ei}, we consider the multivariate linear regression model with Gaussian DAG errors. 
		Our goal in this paper is to jointly recover the true sparsity pattern in the coefficient matrix $B$ and Cholesky factor $L$.
		In this section, we introduce a Bayesian approach based on the exact likelihood.

		\subsection{Priors for $B$, $L$, $D$ and $\calD$}\label{subsec:priors}
		
		For the coefficient matrix $B= (b_{kj})\in\bbR^{p\times q}$, we consider the following spike and slab priors,  for $1\le k\le p$ and $1\le j \le q$, 
		\bean
		\gamma_{kj} &\sim& \mbox{Bernoulli}(\eta_1), \label{model:gamma} \\
		b_{kj} \mid  \gamma_{kj} = 0  &\sim& \delta_0 , \label{model:beta_0}\\
		b_{kj} \mid  \gamma_{kj} = 1  &\sim& N(0, \tau_1^2), \label{model:beta_1}
		\eean
		for some constants $0<\eta_1<1$ and $\tau_1>0$, where $\delta_0$ is a point mass at zero.
		In the above, $\gamma_{kj}$ is a  binary latent variable such that $\gamma_{kj} = 1$ if and only if  $b_{kj}\neq 0$.
		Let $\Gamma = (\gamma_{kj}) = (\gamma_1, \ldots, \gamma_q) = (\Gamma_1, \ldots, \Gamma_p)^T \in \{0,1\}^{p \times q}$ be the collection of all variable indicators. 
		Prior \eqref{model:gamma} can be seen as a collection of independent Bernoulli priors with the inclusion probability $\eta_1$. 		
		In numerical studies, we fixed the prior inclusion probability at $\eta_1=1/p$ as suggested by \cite{Samanta2022}. 
		As an alternative, a beta hyperprior can be used for $\eta_1$ to adaptively select the inclusion probability \citep{castillo2015bayesian}.
		Prior \eqref{model:beta_0} represents the spike part (a point mass at zero) that shrinks insignificant coefficients to zero, while prior \eqref{model:beta_1} represents the slab part that models the nonzero coefficients.

		For $\mathcal{D}$ and $(L,D) \in \Theta_{\mathcal{D}}$, we assume the hierarchical DAG-Wishart prior described in \eqref{model:dag} and \eqref{model:LD} with $\phi_j(\calD) - \nu_j(\calD) = 10$ for $j = 1, \ldots, q$. 
		The conjugate nature of the DAG-Wishart distribution in \eqref{model:LD} will facilitate the derivation of a blocked Gibbs sampler as elaborated in the following section.

		\subsection{Blocked Gibbs sampler}\label{subsec:bGibbs}
		 
		Although the joint posterior distribution is intractable, the conditional posteriors are available in closed form, which enables the use of a blocked Gibbs sampler.
		We have detailed the derivation of the following conditional posteriors in the supplementary material.
		\begin{enumerate}
			\item  Set the initial values for $B, \Gamma, L, D$ and $\calD$.

			\item Run the following steps for $k\in\{1,\ldots, p\}$ and $j\in\{1,\ldots q\}$.
			\begin{enumerate}
				\item Sample $\gamma_{kj} \mid B_{-kj}, \Gamma_{-kj},L,D, \calD, Y \sim \mbox{Bernoulli}( {\nu_{kj} }/ (1+\nu_{kj}) )$,
				where $\nu_{kj} = \eta_1 \{(1-\eta_1) \tau_1 \sqrt{C_{1,kj}} \}^{-1} \exp \{ {C_{2,kj}^2 }/ (2C_{1,kj})  \}$,
				$C_{1,kj}=\omega_{jj} \sum_{i=1}^n x_{ik}^2 +\tau_1^{-2}$ and $C_{2,kj} = \sum_{i=1}^n x_{ik} \big( \sum_{r=1}^q y_{ir} \omega_{jr}- \sum_{l\neq j} b_l^T X_i \omega_{jl} - \omega_{jj}\sum_{s\neq k} b_{sj}x_{is} \big) $.
				\item Sample $b_{kj}\mid \gamma_{kj}, B_{-kj}, \Gamma_{-kj},L,D, \calD, Y  \sim (1-\gamma_{kj})\delta_0 + \gamma_{kj}N ( C_{1,kj}^{-1} C_{2,kj},  \, C_{1,kj}^{-1} )$.
			\end{enumerate}
			
			\item Sample $d_{q}\mid   E \sim \mbox{Inverse-Gamma}\big( (\phi_q(\calD) + n)/2  - 
			1,  n \tilde{s}_{qq} /2 \big )$,  where $\phi_q(\calD)=10$, 
			$\tilde{S} = (\tilde{s}_{jl}) = (nS+U)/n = \{ \sum_{i=1}^n (Y_i - B^TX_i)(Y_i - B^TX_i)^T +U \}/n$ and $E = (E_1,\ldots, E_n)^T$.
			\item Run the following steps for $j \in \{1, \ldots, q-1\}$.
			\begin{enumerate}
				\item Sample $pa_j(\calD)_{new} \sim q\big(\cdot\mid pa_j(\calD)_{old}\big)$, and accept $pa_j(\calD)_{new}$ with the probability
				\bea
				\min \Bigg\{1, \frac{\pi\big(pa_j(\calD)_{new}\mid  E\big) q\big(pa_j(\calD)_{old}\mid pa_j(\calD)_{new}\big)}{\pi\big(pa_j(\calD)_{old}\mid E\big) q\big(pa_j(\calD)_{new}\mid pa_j(\calD)_{old}\big)}\Bigg\} ,
				\eea
				where 
				$\pi(pa_{j}(\calD) \mid E) \propto  {\zeta_{\calD}^j(n\tilde S, \phi  + n) }/\{\zeta_{\calD}^j(U, \phi ) \} \eta_2^{\nu_j(\mathcal{D})} (1-\eta_2)^{q-j-\nu_j(\mathcal{D})}$ and $\zeta_{\calD}^j(U, \phi )$ is the normalizing constant defined in Section \ref{sec2.1}.
				
				\item Sample $d_{j} \mid pa_j(\mathcal D), E \sim \mbox{Inverse-Gamma} (\{\phi_j (\calD)+ n - \nu_{i}(\mathcal D)\} /2  - 
				1, n \tilde S_{j|pa_j({\mathcal{D}})} / 2 )$.
				
				\item Sample $L_{\mathcal D.j}^> \mid d_{j}, pa_j(\mathcal D), E \sim N_{\nu_j(\mathcal D)} (-(\tilde S_{\calD}^{>j})^{-1}\tilde S_{\calD.j},  n^{-1} d_j(\tilde S_{\calD}^{>j})^{-1} )$.
				
			\end{enumerate}
			
			\item Repeat Steps 2 and 4 until a sufficiently long chain is acquired.
			 
		\end{enumerate}

		The kernel $q\big(\cdot\mid pa_j(\calD)_{old}\big)$ in step 4(a) is chosen to form a new set $pa_j(\calD)_{new}$ by randomly deleting an edge from $pa_j(\calD)_{old}$ with probability 0.5 or adding an edge to $pa_j(\calD)_{old}$ with probability 0.5. 
		In the above algorithm, step 4 can be parallelized for each column to enhance computational speed.
		For more details, we refer interested readers to \cite{cao2019posterior} and \cite{lee2019minimax}.

		\subsection{Selection and estimation based on MCMC samples}\label{subsec:est_sel}
		
		We can obtain an estimated variable indicator $\hat \Gamma$  using the median probability model (MPM) \citep{barbieri2004optimal} as follows. 
		Suppose we run the above Gibbs sampler for an appropriate number $M$ of iterations after the burn-in period and obtain the MCMC samples, $\{B^{(i)}, \Gamma^{(i)}\}_{i=1}^M$. 
		The MPM can be constructed as
		\bea 
		\hat \gamma_{kj} =  \begin{cases}
			1, \quad\mbox{if } \frac 1 M \sum_{i = 1}^M \gamma_{kj}^{(i)} \ge \frac 1 2,\\
			0, \quad\mbox{otherwise}.
		\end{cases} 
		\eea
		When the posterior probability of the posterior mode is larger than $0.5$, the MPM corresponds to the posterior mode \citep{barbieri2004optimal}. 
		Similarly, we can obtain an estimated DAG $\hat{\calD}$ using the MPM based on the MCMC samples, $\{\calD^{(i)} \}_{i=1}^M$.
		
		Suppose we have obtained $\hat \gamma_{kj}$'s as above.
		Then we obtain the estimated coefficient matrix $\hat{B}= (\hat{b}_{kj})$ by
		\bean\label{hat_bkj_MCMC}
		\hat b_{kj} &=& \frac{\sum_{i = 1}^Mb_{kj}^{(i)}I\big(\gamma_{kj}^{(i)} = 1\big)}{\sum_{i = 1}^M I\big(\gamma_{kj}^{(i)} = 1\big)} \,\, \text{for $(k,j)$ such that $\hat \gamma_{kj}=1$}
		\eean
		and $\hat{b}_{kj}=0$ for $(k,j)$ such that $\hat{\gamma}_{kj}=0$.
		For  given estimated DAG $\hat{\calD}$ and coefficient matrix $\hat{B}$, finally we calculate the estimated Cholesky factor and diagonal matrix, say $\hat{L} =(\hat{L}_{jl})$ and $\hat{D}= \mbox{Diag}(\hat{d}_1,\ldots, \hat{d}_q)$, respectively.
		Specifically, we define $\hat{L}_{jl}$ as a similar manner to \eqref{hat_bkj_MCMC} for $(j,l)\in \hat{\calD}$ and $\hat{L}_{jl}=0$ otherwise.
		Alternatively, the posterior means can be used, i.e., we can use $\hat{L}^>_{\hat{\mathcal D} .j } = E(L^>_{{\mathcal D} .j } \mid d_j, pa_j(\hat{\mathcal D}) , \hat{E}) = E(L^>_{{\mathcal D} .j } \mid pa_j(\hat{\mathcal D}) ,\hat{E})$, where $\hat{E}$ represents the estimated error with $B$ replaced by $\hat{B}$.
		The exact form of the posterior mean is given in the supplementary material.
		To obtain $\hat{D}=\mbox{Diag}(\hat{d}_1,\ldots, \hat{d}_q)$, we use the posterior mode $\hat{d}_{j , \hat{\mathcal{D}} }= \argmax_{d_j} \pi(d_j \mid pa_j(\hat{\mathcal{D} }) , \hat{E} )$, whose exact form is given in the supplementary material.

		\subsection{Computational complexity of the exact likelihood-based approach}\label{subsec:comp_exact}
		
		In this section, we describe the computational complexity per iteration of the MCMC algorithm in Section \ref{subsec:bGibbs}. 
		Let 
		$M = B^TX^TX \in \bbR^{q\times p}$.
		Within the for loops in step 2, we need to compute $(M_{.k})^T\Omega_{.j}$ and update the $j$th row of $M$, for $k=1,\ldots,p$ and $j=1,\ldots, q$. 
		These operations require computational costs of $O(q)$ and $O(p)$, respectively.		
		To sum up, the total number of operations for sampling $B$ and $\Gamma$ is $O\big(pq(p \vee q)\big)$. 
		Next, computing $S= n^{-1}\sum_{i=1}^n (Y_i - B^TX_i)(Y_i - B^TX_i)^T $ has a complexity of $O\big(nq(p\vee q)\big)$
		and updating $\Omega = LD^{-1}L^T$ in the end of each iteration yields at most $O(q^3)$ operations (the sparsity in $L$ can lessen the complexity). 
		The other steps of sampling $L$ and $D$ require $O\big(\nu_j(\mathcal{D})^3\big)$ operations for each $j = 1, \ldots, q-1$, although $\nu_j(\mathcal{D})$ may vary with each iteration.
		
		Hence, under a high-dimensional setting where $n<\min(p,q)$, the total cost of computational complexity for each MCMC iteration of the exact algorithm is
		$$O\big(pq(p \vee q )\big) +  O(q^3) + O\Big(\sum_{j = 1}^{q-1}\nu_j(\mathcal{D})^3\Big).$$
		This is comparable to the JRNS algorithm in \cite{Samanta2022}, which, however, is based on the generalized likelihood instead of the exact likelihood and does not guarantee the positive definiteness of the precision matrix.
		Note that the above computational cost can be substantial in high-dimensional settings with large $p$ and $q$.

		\section{A two-step approach for scalable computation}  \label{sec:two-step}

		In this section, we further develop a two-step approach for scalable inference by leveraging the step-wise approach proposed by \cite{Samanta2022}.		
		In the first step, we ignore the dependency structure among $q$ response variables in \eqref{model:Yi} and adopt a generalized likelihood. 
		Based on this, we estimate the regression coefficient vectors for $q$ univariate linear regression models. 
		In the second step, we utilize the estimated coefficient matrix $\hat{B}$ obtained from the first step to obtain the {\it estimated error} matrix $\hat{E} = Y - X \hat{B}$. 
		By assuming this as a random sample from a Gaussian DAG, we estimate $(L, D, \mathcal{D})$.
		Now, we will sequentially describe in detail the two-step approach that we propose.

		\subsection{First step: estimating $\Gamma$ and $B$}
		The marginal model for the $j$th response variable derived from model \eqref{model:Yi} is the following univariate linear regression model:
		\bea
		y_j = Xb_j + \epsilon_j, \quad j=1,\ldots, q, 
		\eea 
		where $y_j \in\bbR^n$, $X\in\bbR^{n\times p}$, $b_j\in\bbR^p$ and  $\epsilon_j \sim N_n(0, \sigma_j^2 I_n )$.
		Let $\sigma_j^2$ be the $j$th diagonal entry of $\sg$. 		
		Building on this, we propose using the product of the likelihoods from $q$ univariate linear regression models as an approximate likelihood for the model:
		\bea
		L_{g}\big(Y \mid  B\big) \,:=\, \prod_{j=1}^q (2\pi \sigma_j^2)^{-\frac n 2}\exp\Big(-\frac 1 {2\sigma_j^2} \| y_j - Xb_j \|_2^2\Big) ,
		\eea
		where $\|\cdot\|_2$ denotes the vector $\ell_2$-norm.
		We refer to this as the generalized likelihood.
		Note that the $q$ response variables may be dependent on each other. 
		However, the above generalized likelihood assumes independence by setting $\Sigma = \mbox{Diag}(\sigma_1^2,\ldots, \sigma_q^2)$. 
		This is a key feature that enables the introduction of the empirical sparse prior, which will be described shortly.
		To define the empirical sparse prior, we require a (sequence of) univariate regression models.  
		However, in the exact approach presented in Section \ref{sec:model}, such an interpretation in terms of univariate regressions is not feasible.  
		For this reason, the empirical sparse prior is employed only in the two-step approach.

		\subsubsection{Empirical sparse prior}
		
		To assign priors to the parameters $(b_j, \sigma_j^2)$ of the generalized likelihood, we adapt the empirical sparse Cholesky prior proposed by \cite{lee2019minimax} to suit our setting.		
		Let $|\gamma_j| = \sum_{k = 1}^p \gamma_{kj}$, $b_{j, \gamma_j} = (b_{lj})^T_{l:\gamma_{lj} \neq 0} \in \bbR^{|\gamma_j|}$ and $X_{\gamma_j} \in \bbR^{n \times |\gamma_j|}$ be the number of nonzero entries in $b_j$, the vector consisting of nonzero entries in $b_j$ and the submatrix of $X$ consisting of active columns indexed in $\gamma_j$, respectively.
		The proposed prior is as follows, and we refer to it as the empirical sparse prior:
		\bean
		\pi(\gamma_j) \,\,&\propto& \,\, \binom{p}{|\gamma_j|}^{-1} p^{- c_1 |\gamma_j|} I( 0\le |\gamma_j| \le R_j ) , \label{gamma_j} \\
		\pi(\sigma_j^2) \,\,&\propto&\,\,   (\sigma_j^2)^{ - \nu_0/2 -1 } , \label{sigma_j}\\
		b_{j , \gamma_j} \mid \sigma_j^2, \gamma_j  \,\,&\sim&\,\,  N_{|\gamma_j|} \Big(  \hat{b}_{j, \gamma_j}  , \,\, \frac{\sigma_j^2}{\kappa} \big( X_{\gamma_j}^T X_{\gamma_j}  \big)^{-1}  \Big)   , \label{b_j} 
		\eean
		for some positive constants $\kappa$, $\nu_0$ and a positive integer $0< R_j \le p$, where $\hat{b}_{j, \gamma_j} = ( X_{\gamma_j}^T X_{\gamma_j} )^{-1} X_{\gamma_j}^T y_j$.

		The prior $\pi(\gamma_j)$ in \eqref{gamma_j} acts as a penalty for large models and reflects the prior knowledge that the true model is sparse.
		We adopt the improper prior for $\sigma_j^2$ in \eqref{sigma_j} for computational convenience, but it can be generalized to a proper inverse-gamma prior, and the following analysis still holds. 
		Note that the conditional prior for $b_{j , \gamma_j}$ in \eqref{b_j} corresponds to the Zellner's $g$-prior \citep{zellner1986assessing} centered at the least square estimator $ \hat{b}_{j, \gamma_j}$. 
		The term {\it empirical} in empirical sparse prior refers to the fact that the prior for $b_{j,\gamma_j}$ is data-dependent.
		Similar priors have been used in \cite{martin2017empirical} and \cite{lee2019minimax}.

		
		\begin{remark} 
			In this two-step approach, we use the empirical sparse prior \eqref{gamma_j}--\eqref{b_j} instead of the ``standard'' spike and slab prior $b_{kj} \mid \sigma_j^2 \sim (1 - \eta_1)\delta_0 + \eta_1 N(0, \sigma_j^2/\kappa)$ which is similar to the one used in Section \ref{subsec:priors}.
			However, in fact, this standard spike and slab prior can also be applied within the two-step framework. 
			Moreover, as investigated later in the supplementary material, the empirical sparse prior and the standard spike and slab prior yield similar performance in most settings.
			The primary motivation for using the empirical sparse prior is to facilitate the theoretical analysis. 
			Combined with the $\alpha$-fractional likelihood, this prior allows us to establish selection consistency for $\Gamma$ and posterior convergence rates for $B$.
		\end{remark}

		\subsubsection{$\alpha$-fractional posterior}
		
		We use the fractional likelihood of $L_g(Y\mid B)$ with the power $\alpha \in (0,1)$ to calculate the posterior, which is defined as
		\bea
		L_\alpha (Y \mid  B) \,\,:=\,\, L_{g}\big(Y \mid  B\big)^\alpha 
		\,\,\equiv\,\, \prod_{j=1}^q L_\alpha ( b_{j , \gamma_j}, \sigma_j^2, \gamma_j) , 
		\eea
		where $L_\alpha ( b_{j , \gamma_j}, \sigma_j^2, \gamma_j) = (2 \pi\sigma_j^2)^{- \alpha n/2} $ $\exp  \{-\alpha / (2 \sigma_j^2) \| y_j - X_{\gamma_j}b_{j,\gamma_j}\|_2^2 \}$ for $j = 1, \ldots, q$.	
		There are two main reasons for using the fractional likelihood. 
		First, it facilitates the proof of theoretical properties of the resulting posterior under relatively weaker conditions compared to the exact posterior \citep{bhattacharya2019bayesian}. 
		Second, it helps prevent the potential issue of the posterior tracking the data too closely, which can arise because the prior \eqref{b_j} is data-dependent \citep{martin2017empirical,lee2019minimax}.
		
		Under the above fractional likelihood and the empirical sparse prior \eqref{gamma_j}--\eqref{b_j}, we obtain the following conditional posterior distributions:
		\bea
		\pi_\alpha(\gamma_j \mid  y_j) &\propto& \pi(\gamma_j)\Big(1 + \frac{\alpha}{\kappa} \Big)^{- \frac{|\gamma_j|}{2}}  (\hat{\sigma}_{j, \gamma_j}^2)^{-\frac{\alpha n + \nu_0}{2} },\\
		\sigma_j^2 \mid \gamma_j ,  y_j &\sim&  \mbox{Inverse-Gamma} \Big(  \frac{\alpha n + \nu_0}{2} , \,\, \frac{\alpha n}{2} \hat{\sigma}_{j, \gamma_j}^2 \Big), \\
		b_{j , \gamma_j} \mid  \sigma_j^2, \gamma_j,  y_j &\sim& N_{|\gamma_j|} \Big( \hat{b}_{j, \gamma_j},\,\,  \frac{\sigma_j^2}{\alpha + \kappa} \big( X_{\gamma_j}^T X_{\gamma_j} \big)^{-1} \Big),  
		\eea
		where $ \hat{\sigma}_{j, \gamma_j}^2  = n^{-1} y_j^T(I_{n} - P_{\gamma_j}) y_j$ and $P_{\gamma_j} = X_{\gamma_j}( X_{\gamma_j}^T X_{\gamma_j})^{-1} X_{\gamma_j}^T$,  for $j=1,\ldots, q$.
		We denote the posterior by $\pi_\alpha(\gamma_j \mid  y_j)$ and refer to it as the $\alpha$-fractional posterior, to emphasize that the $\alpha$-fractional likelihood is used.

		\subsubsection{Posterior inference for $\Gamma$ and $B$}
		For posterior computation, we run the following Metropolis-Hastings (MH) algorithm to draw $\gamma_j$'s from $\pi_\alpha(\gamma_j \mid  y_j)$ for each $j=1,\ldots, q$.
		By using the posterior samples obtained in the below MH algorithm, we can obtain the estimated variable indicator $\hat{\Gamma}$ through the MPM previously described in Section \ref{subsec:est_sel}.
		We use the same kernel as in Section \ref{subsec:bGibbs}, but replace $pa_j(\mathcal{D})$ with $\gamma_j$.
		
		\begin{enumerate}
			\item  Set the initial value for $\gamma_j$.
			\item Sample $\gamma_j^{new} \sim q\big(\cdot\mid \gamma_j^{old}\big)$, and accept $\gamma_{j}^{new}$ with the probability
			\bea
			\min \Bigg\{1, \frac{\pi_\alpha(\gamma_j^{new} \mid  y_j)q(\gamma_j^{old}\mid \gamma_j^{new})}{\pi_\alpha(\gamma_j^{old} \mid  y_j)q(\gamma_j^{new}\mid \gamma_j^{old})}\Bigg\} .
			\eea
			
			\item Repeat Step 2 until a sufficiently long chain is acquired.
		\end{enumerate}

		
		Given $\hat{\Gamma}$, an estimated coefficient matrix $\hat B$ can be constructed without sampling $b_j$'s by aligning the posterior mean $\hat{b}_{j, \hat \gamma_j} = (X_{\hat{\gamma}_j}^T X_{\hat{\gamma}_j} )^{-1} X_{\hat{\gamma}_j}^T y_j$ for $j = 1, \ldots, q$.
		Alternatively, if the uncertainty of $B$ is of interest, it is possible to obtain posterior samples for $(b_{j,\gamma_j}, \sigma_j^2)$ using the above conditional posterior given $\gamma_j$.

		\subsection{Second step: estimating $\mathcal D$ and $(L, D)$}
		
		Let $\hat{B}$ be the estimated coefficient matrix based on the posterior mean $\hat{b}_{j, \hat \gamma_j}$ obtained by the first step.
		Denote $\hat E = (\hat E_1, \ldots, \hat E_n)^T$ with $\hat E_i = y_i - \hat B^T X_i$ for $i = 1, \ldots, n$ as  the resulting error estimates. 
		Note that the actual errors $E_i$'s are random samples following $N_q(0, (L D^{-1} L^T)^{-1})$. 
		In the second step, we will adopt an approximate model that considers the estimated errors $\hat{E}_i$'s as random samples following $N_q(0, (L D^{-1} L^T)^{-1})$, i.e., 
		\bean\label{model:hatEi}
		\hat{E}_i \mid L, D, \calD &\overset{iid}{\sim}& N_q(0, (L D^{-1} L^T)^{-1}) ,\quad i=1,\ldots, n.
		\eean
		Then, our task of inferring the dependency structure of response variables in multivariate linear regression models boils down to the problem of estimating a sparse Cholesky factor $L \in \calL_{\calD}$ and a DAG $\calD$ in \eqref{model:hatEi}.

		For the inference of $(L, D)$ and $\mathcal{D}$ in the above model, we utilize the hierarchical DAG-Wishart prior \eqref{model:dag} and \eqref{model:LD} considered earlier.
		Through straightforward calculations, it can be verified that the posterior distributions of $(L, D)$ and $\mathcal{D}$ are identical to those from the exact likelihood-based approach detailed in the supplementary material, with the exception that $S= n^{-1}\sum_{i=1}^n (Y_i - B^TX_i)(Y_i - B^TX_i)^T $ is replaced by $\hat{S} = n^{-1} \sum_{i=1}^n (Y_i - \hat{B}^T X_i) (Y_i - \hat{B}^T X_i)^T$.

		For efficient posterior inference, we first obtain posterior samples of $\mathcal{D}$ and then proceed with DAG selection based on these samples. 
		To obtain posterior samples of $\mathcal{D}$, we use the MH algorithm in step 4 of Section \ref{subsec:bGibbs}, replacing $E$ with $\hat{E}$.
		Subsequently, when estimating $(L, D)$, we can perform posterior inference without MCMC by leveraging $\pi(L, D \mid \hat{\mathcal{D}}, \hat{E})$, where $\hat{\mathcal{D}}$ is the estimated DAG. 
		For instance, point estimates for $(L, D)$ can be calculated in a closed form, such as posterior mean or mode. 
		Efficient inference of this kind is possible due to the two-step approach, contributing to the reduction in computational burden. 
		In the exact likelihood-based method, $(B, \Gamma)$ are always included in  conditional posteriors for other parameters, making it impossible to separately infer $\mathcal{D}$ using the MH algorithm and infer $(L, D)$ without an MCMC algorithm.
		Of course, if there are no time constraints, Metropolis within Gibbs sampler can be used to obtain MCMC samples for $(L, D, \mathcal{D})$, enabling full posterior inference.


		\subsection{Computational complexity of the two-step approach}\label{subsec:comp_two}

		When sampling $\gamma_j$ from $\pi(\gamma_j\mid y_j)$, the main computational bottleneck is to compute $\hat{\sigma}_{j,\gamma_j}^2$. By precomputing $Y^T Y$, $X^T Y$ and $X^T X$ before the algorithm starts, its computational cost is $O(|\gamma_j|^3)$ for $j=1,\ldots, q$, leading to a total of $O\big(\sum_{j = 1}^q|\gamma_j|^3\big)$ operations.
		If necessary, $(b_{j,\gamma_j}, \sigma_j^2)$ can be sampled from the conditional posterior, and their computational cost is also $O(|\gamma_j|^3)$.
		After obtaining the posterior samples of $\gamma_j$'s, calculating $\hat{\Gamma}, \hat{B}, \hat{E}$ and $\hat{S}$ is performed only once outside of the MCMC iterations, and therefore, it is negligible in the overall computational cost.		
		Next, the remaining steps of generating $pa_j(\mathcal{D})$ also need a complexity $O\Big(\sum_{j = 1}^{q-1}\nu_j(\mathcal{D})^3\Big)$, the same as in the exact algorithm.
		
		To sum up, the overall computational cost for each iteration of the two-step approach is
		$$O\Big(\sum_{j = 1}^q|\gamma_j|^3\Big) + O\Big(\sum_{j = 1}^{q-1}\nu_j(\mathcal{D})^3\Big).$$		 
		This is a significantly smaller value compared to the exact algorithm, especially in sparse high-dimensional settings where $|\gamma_j| \ll p^{2/3}$ and $\nu_j(\mathcal{D}) \ll q^{2/3}$ for all $j$. 
		Moreover, in this sparse regime, this results in lower computational complexity even compared to the stepwise method of \cite{Samanta2022}, which requires $O(p^2q + q^3)$.

		\subsection{Theoretical properties of the two-step approach}\label{sec:theory}
		
		In this section, we present asymptotic properties of the posterior in the two-step approach.
		Let $B_0 =(b_{0,jl})$ and $\Gamma_0=(\gamma_{01},\ldots, \gamma_{0q})$ be the true coefficient matrix  and its support, respectively.  
		Similarly, let $\sg_0= (L_0 D_0^{-1} L_0^T)^{-1} $ and $\calD_0$ be the true covariance matrix and DAG, respectively. 
		The main result is twofold:
		(1) strong selection consistency and posterior convergence rates for the coefficient matrix $B_0$
		and
		(2) posterior ratio consistency and strong selection consistency for the Cholesky factor $L_0$.
		The proofs of the main theoretical results are provided in the supplementary material.

		\subsubsection{Selection consistency and posterior convergence rates for $B_0$}\label{subsec:theory_B0}
		
		We first establish the results for $B_0$. 
		The following conditions are sufficient to achieve strong selection consistency and posterior convergence rates for $B_0$, with a detailed discussion of each condition provided in the supplementary material.
		
		\begin{itemize}
			\item[(A1)] (dimension) $\max(n, q) \le p$ and $\log q = o(n)$.
			\vspace{-.2cm}
			
			\item[(A2)] (coefficient matrix $B_0$) For some $C_{\rm bm} > c_2 +2$ and small $0<\epsilon_0<1/2$, 
			\bea
			\min_{(j,l): b_{0,jl} \neq 0 } \frac{b_{0,jl}^2}{\sigma_{0j}^2 }  &\ge&   \frac{16}{\alpha(1-\alpha) \epsilon_0 (1-2\epsilon_0)^2 } C_{\rm bm} \frac{\log p }{n}   ,
			\eea
			$\sigma_{0j}^{-2} \| X_{\gamma_{0j}} b_{0j, \gamma_{0j} } \|_2^2 \le 2n(1+\epsilon_0)^2 \epsilon_0^{-2}  $ and $|\gamma_{0j}| \le R_j$ for all $j=1,\ldots, q$. 
			\vspace{-.2cm}
			
			\item[(A3)] (design matrix)
			For some small $0<\epsilon_0<1/2$, 
			\bea
			(1-2\epsilon_0)^2 \epsilon_0 \,\,\le\,\, \Phi_{\min} \,\,\le\,\, \Phi_{\max} \,\,\le\,\, (1+2\epsilon_0)^2 \epsilon_0^{-1} ,
			\eea
			where 
			\bea
			\Phi_{\min} &=& \min_{\gamma_j: \gamma_j \nsupseteq \gamma_{0j}, |\gamma_j|\le R_j } \lambda_{\min} \big( n^{-1}  X_{\gamma_{0j} \cup \gamma_j }^T X_{\gamma_{0j} \cup \gamma_j } \big) , \\
			\Phi_{\max} &=& \max_{\gamma_j: \gamma_j \nsupseteq \gamma_{0j}, |\gamma_j|\le R_j } \lambda_{\max} \big( n^{-1}  X_{\gamma_{0j} \cup \gamma_j }^T X_{\gamma_{0j} \cup \gamma_j } \big) .
			\eea
			\vspace{-.6cm}
			
			\item[(A4)] (hyperparameters) $\nu_0=o(n)$, $\kappa=O(1)$, $0<\alpha<1$, $c_1 \ge 2$ and $R_j = \lfloor c_3 n / \log p  \rfloor$, 
			where $c_3=(\epsilon')^2 \epsilon_0^2 /\{128 (1+2\epsilon_0)^2 \}$ and $\epsilon' = \{ (1-\alpha) /10 \}^2$.
			\vspace{-.2cm}
			
		\end{itemize}

		Now we state strong selection consistency for the coefficient matrix $B_0$ (Theorem \ref{thm:selection_gamma}).
		It says that the posterior probability for the true model $\Gamma_0$, which is the support of $B_0$, converges to one in probability as we observe more data.

		\begin{theorem}[Selection consistency for $B_0$]\label{thm:selection_gamma}
			Under  conditions (A1)--(A4), we have 
			\bea
			\bbE_0 \Big\{  \pi_\alpha \big(  \Gamma \neq \Gamma_{0} \mid Y \big)  \Big\} &=& o(1)  \quad \text{as } n\to \infty .
			\eea
		\end{theorem}

		Theorem \ref{thm:post_conv_B} provides posterior convergence rates for each column of $B_0$ under vector $\ell_1$- and $\ell_2$-norms.
		
		\begin{theorem}[Posterior convergence rates for $B_0$]\label{thm:post_conv_B}
			Let $s_0 = \max_{1\le j\le q} |\gamma_{0j}| $. 
			Under  conditions (A1)--(A4) and $\max(s_0 , \log p ) = o(n)$, we have 
			\bea
			\bbE_0 \Big\{ \pi_\alpha \Big(  \cup_{1\le j\le q} \Big\{  \|b_j - b_{0j}\|_1 \ge K \sigma_{0j} \Big(  \frac{ |\gamma_{0j}|^2+ |\gamma_{0j}|\log p  }{n}  \Big)^{1/2}  \Big\} \mid Y \Big)  \Big\}  &=& o(1)   , \\
			\bbE_0 \Big\{ \pi_\alpha \Big(  \cup_{1\le j\le q} \Big\{  \|b_j - b_{0j}\|_2 \ge K \sigma_{0j}  \Big(  \frac{ |\gamma_{0j}|+ \log p }{n}  \Big)^{1/2}  \Big\} \mid Y \Big)  \Big\}  &=& o(1)   ,
			\eea
			as $n\to\infty $, for some constant $K \equiv K_{\alpha,\kappa,\epsilon_0}>0$ depending only on $(\alpha, \kappa, \epsilon_0)$.
		\end{theorem}

		In the Bayesian literature, \cite{ning2020bayesian} and \cite{Samanta2022} also established selection consistency and posterior convergence rates for $B_0$ but under slightly different models and conditions. 
		We provide a detailed comparison with these works in the supplementary material.

		\subsubsection{Selection consistency for $L_0$}\label{subsec:theory_L0}
		
		Next, we present the results related to selection properties for the Cholesky factor $L_0$ (Theorems \ref{thm:post_ratio_L} and \ref{thm:strong_sel_L}).
		To establish theoretical properties of the posterior, we restrict the support of prior \eqref{model:dag} so that $\nu_j(\calD) \le \xi_j$ for $j=1,\ldots, q-1$.
		It ensures that we do not consider unreasonably large models.
		We further assume the following conditions, with a detailed discussion of each provided in the supplementary material.		
		\vspace{-.2cm}
		\begin{itemize}
			
			\item[(A5)] (bounded eigenvalues) There exists $0<\epsilon_{0,n}<1$ such that $\epsilon_{0,n} \le \lambda_{\min}(\sg_0) \le \lambda_{\max}(\sg_0) \le \epsilon_{0,n}^{-1}$.
			\vspace{-.2cm}
			
			\item[(A6)] (model size)  $ \epsilon_{0,n}^{-8} \xi_0^2 s_0  ( s_0+\log p)  = o(n)$, where $\max_j \xi_j \le \xi_0$.
			\vspace{-.2cm}

			\item[(A7)] (Cholesky factor $L_0$) $\max_j \nu_j(\mathcal{D}_0)/\xi_j \le 1$ and, for some $\eta_n$ defined at condition (A8),  
			\bea
			\min_{(l,j): L_{0,lj} \neq 0 }L_{0,lj}^2 &\ge& 40 \eta_n \epsilon_{0,n}^{-2} . 
			\eea 
			\vspace{-.7cm}
			
			\item[(A8)] (hyperparameters) There exist global constants $c, \delta_1$ and $\delta_2$ such that 
			$\phi_j(\calD) - \nu_j(\calD)= c > 2$ for every $\calD$ and $1\le j\le p$, 
			and $0<\delta_1 \le \lambda_{\min}(U) \le \lambda_{\max}(U) \le \delta_2 <\infty$.
			Furthermore, we set to $\eta_2  = \exp (- \eta_n n)$ for some sequence $\eta_n$ satisfying $\epsilon_{0,n}^{-4} \xi_0 \sqrt{s_0(s_0+\log p)/n } = o(\eta_n)$.
			\vspace{-.2cm}
			
		\end{itemize}

		We first state posterior ratio consistency for $L_0$ based on the two-step approach, which means the posterior probability of the true graph $\calD_0$ is much larger than those of other graphs with probability tending to one.
		
		\begin{theorem}[Posterior ratio consistency for $L_0$]\label{thm:post_ratio_L}
			Suppose the conditions in Theorem \ref{thm:post_conv_B} hold.
			If we further assume that conditions (A5)--(A8), we have 
			\bea
			\max_{\calD \neq \calD_0} \frac{\pi(\calD \mid \hat{E}) }{\pi (\calD_0 \mid \hat{E}) } \overset{p}{\lra} 0   \quad \text{as } n\to \infty .
			\eea
		\end{theorem}

		Theorem \ref{thm:post_ratio_L} implies that  the true graph $\calD_0$ coincides with the maximum a posteriori (MAP) estimator $\hat{\calD} = \argmax_{\calD} \pi(\calD \mid \hat{E})$ with probability tending to one.
		However, it does not guarantee that the posterior probability of $\calD_0$ is sufficiently large.
		For example, even though Theorem \ref{thm:post_ratio_L} holds, $\pi(\calD_0\mid \hat{E})$ might converge to zero with probability tending to one.
		The next theorem presents a stronger result, which says that the posterior probability of $\calD_0$ converges to one in probability.

		\begin{theorem}[Selection consistency for $L_0$]\label{thm:strong_sel_L}
			Under conditions in Theorem \ref{thm:post_ratio_L},  
			\bea
			\bbE_0 \Big\{ \pi(\calD \neq \calD_0 \mid \hat{E})  \Big\}
			&=&  o(1)   \quad \text{as } n\to \infty .
			\eea
		\end{theorem}

		The proofs of Theorems \ref{thm:post_ratio_L} and \ref{thm:strong_sel_L} largely use techniques from \cite{cao2019posterior}. 
		The primary difference lies in the fact that the ``data'' used to calculate the posterior of $\mathcal{D}$  is the estimated error $\hat{E}_i$ in this paper, whereas it is the true error $E_i$ in \cite{cao2019posterior}.
		To address this, we have demonstrated that $\hat{E}_i$ is sufficiently close to $E_i$ with high probability. 
		Specifically, we have proved that $\hat{S} = n^{-1}\sum_{i=1}^n \hat{E}_i \hat{E}_i^T$ is sufficiently close to $n^{-1}\sum_{i=1}^n E_i E_i^T$ with high probability tending to one. 
		Once this is guaranteed, the remainder of the proof can be derived by appropriately modifying the bounds used in \cite{cao2019posterior}.

		\section{Simulation Studies} \label{sec:simul}
		
		In this section, we evaluate the performance of proposed methods under various simulation settings.
		The Gibbs sampler based on the exact likelihood and spike and slab priors in Section \ref{sec:model} referred to as ``ESS'',  and the two-step approach under the empirical sparse priors in Section \ref{sec:two-step} referred to as ``TES''. 
		
		We closely follow but slightly modify the simulation settings in \cite{Samanta2022}. 
		The data is simulated from 
		$Y = XB_0 +E$, 
		where  $X_i  \overset{iid}{\sim} N_p(0, C_0)$ and $E_i \overset{iid}{\sim} N_q(0, \Sigma_0)$ for $i = 1, \ldots, n$, with $C_0 = \big(0.6^{|r-s|}\big)_{r,s = 1}^p$ and $\Sigma_0 =  (L_0D_0^{-1} L_0^T )^{-1}$. 
		We sample the nonzero off-diagonal entries of $L_0$ from $\mbox{Unif}([-0.7, -0.3] \cup [0.3, 0.7])$ and the diagonal entries of $D_0$ from $\mbox{Unif}(2,5)$. 
		The nonzero entries of $B_0$ are chosen according to the following four settings to include different combinations of small and large signals.
		\begin{itemize}
			\vspace{-.35cm}
			\item Setting 1: Generate  the nonzero entries in $B_0$ from $\mbox{Unif}(1.5, 3)$. 
			\vspace{-.2cm}
			\item  Setting 2: Generate  the nonzero entries in $B_0$ from $\mbox{Unif}\big([-3, -1.5] \cup [1.5, 3]\big)$. 
			\vspace{-.2cm}
			\item Setting 3: Generate  the nonzero entries in $B_0$  from $\mbox{Unif}\big([-1.5, -0.5] \cup [0.5, 1.5]\big)$. 
			\vspace{-.2cm}
			\item Setting 4: Generate  the nonzero entries in $B_0$ from $\mbox{Unif}\big([-1.5, -0.5] \cup [1.5, 3]\big)$. 
			\vspace{-.35cm}
		\end{itemize}

		We consider the following scenarios of the triplet $(n, p, q)$ and the sparsity levels in $B_0$ and $L_0$ encoded in $\Gamma_0=(\gamma_{0,kj})$ and $\mathcal D_0$, respectively.
		Here, $\|\Gamma_0\|_0$ denotes the number of nonzero entries in $\Gamma_0$.
		\begin{itemize}
			\vspace{-.35cm}
			\item Scenario 1: $(n, p, q) = (100, 100, 50)$, $\|\Gamma_0\|_0= p/5$ and $\Sigma_0 =  (L_0D_0^{-1} L_0^T )^{-1}$, where $\sum_{j = 1}^q \nu_j(\mathcal D_0) = q/5$.
			\vspace{-.2cm}
			\item Scenario 2: $(n, p, q) = (100, 200, 200)$, $\|\Gamma_0\|_0= p/5$ and $\Sigma_0 =  (L_0D_0^{-1} L_0^T )^{-1}$, where $\sum_{j = 1}^q \nu_j(\mathcal D_0) = q/5$.
			\vspace{-.2cm}
			\item Scenario 3: $(n, p, q) = (150, 300, 200)$, $\|\Gamma_0\|_0 = p/30$ and $\Sigma_0 =  (L_0D_0^{-1} L_0^T )^{-1}$, where $\sum_{j = 1}^q \nu_j(\mathcal D_0) = q/20$.			
			\vspace{-.2cm}
			\item  Scenario 4: $(n,p,q)=(100, 150, 100)$, $\|\Gamma_0\|_0= p/10$ and $\sg_0 = \tilde{\sg}_0 + \{0.01 - \lambda_{\min}(\tilde{\sg}_0)\} I_p$, where  
			$(\tilde{\sg}_0)_{ij} = 2  (  1 - |i-j|/10 ) I(|i-j| \le 5)$.   
			Scenario 4 will show the performance of each method when the network structure of the error is an undirected graph.
			Furthermore, to consider the misspecified ordering case, we randomly shuffle the columns to construct $E$.
			\vspace{-.2cm}
		\end{itemize}


		Lastly, we consider Scenario 5 to assess the variable selection performance when extremely small signals are present, as well as to evaluate the ability to quantify uncertainty through posterior credible intervals.
		In this scenario, we set $(n, p, q) = (100, 100, 50)$, $\|\Gamma_0\|_0= 20$ and $\sum_{j = 1}^q \nu_j(\mathcal D_0) = q/5$. 
		Five of the nonzero entries of $B_0$ were fixed at 1.5, another five at 1, and the remaining ten were generated from $\text{Unif}(0, 0.3)$. 
		The last component represents extremely small signals.

		The proposed methods, ESS and TES, will be compared with the following methods: 
		(i) JRNS and (ii) JRNS.S corresponding to the Joint Regression Network Selector and the stepwise approach proposed in \cite{Samanta2022}, and (iii) DPE (Spike-and-slab lasso with dynamic posterior exploration) and (iv) DCPE (Spike-and-slab lasso with dynamic conditional posterior exploration) proposed in \cite{Deshpande2019}, where the latter two are essentially penalized likelihood estimators obtained by placing $\ell_1$ penalties for individual entries of $B$ and $\Omega$. 
		
		All the tuning parameters in JRNS, JRNS.S, DPE and DCPE are set according to the default values suggested in their GitHub repositories.
		For ESS, as discussed by \cite{Samanta2022} and \cite{cao2019posterior}, we suggest using the following hyperparameters in \eqref{model:dag}--\eqref{model:beta_1}: $\eta_1 = 1/p, \eta_2 = 1/q, \tau_1^2 = 1$, $U = I_q$ and $\phi_j(\calD) - \nu_j(\calD) = 10$ for $j = 1, \ldots, q$. 
		For TES, we set $\alpha = 0.999$ to mimic the Bayesian model with the original likelihood as suggested in \cite{lee2019minimax}.
		We set $R_j = p$ to ensure a fair comparison with other methods.  
		The other hyperparameters were chosen as $\kappa = 0.1$, $\nu_0 = 0$ and $c_1 = 2$ to satisfy the theoretical conditions.
		The initial states for both $L$ and $D$ were set as $I_q$, while the initial state for $B$ and $\Gamma$ was chosen by Lasso  \citep{tibshirani1996regression}. 
		For posterior inference, 2,000 posterior samples were drawn after a burn-in period of 1,000. 
		As the final models for variable and graph selection, we use the MPM \citep{barbieri2004optimal}. 

		The model selection performance of all methods for estimating the sparsity pattern in $B$ is  compared using 
		precision, sensitivity, specificity and Mathews correlation coefficient (MCC) (average over $10$ repetitions). 
		Precision, sensitivity and specificity  are defined as 
		$\mbox{Precision} = {\text{TP}}/ \text{(TP + FP)}$,  $\mbox{Sensitivity} = {\text{TP}}/\text{(TP + FN)}$ and $\mbox{Specificity} = {\text{TN}}/\text{(TN + FP)}$, respectively, 
		where TP, TN, FP and FN correspond to true positive, true negative, false positive and false negative, respectively. 
		Higher values for precision, sensitivity and specificity, closer to 1, suggest better performance.
		MCC is commonly used to assess the performance of binary classification methods and is defined as 
		$$ \text{MCC} = \frac{\text{TP} \times \text{TN} - \text{FP} \times \text{FN}} {\sqrt{\text{(FP + TN)} \times \text{(TP+FN)} \times\text{(TN+FP)}\times\text{(TN+FN)}}} .$$ 
		Note that the value of MCC always fall within the range of $-1$ to 1, where larger values indicate better results.

		\begin{figure}[!tb]
			\centering
			\includegraphics[width=1.0\linewidth]{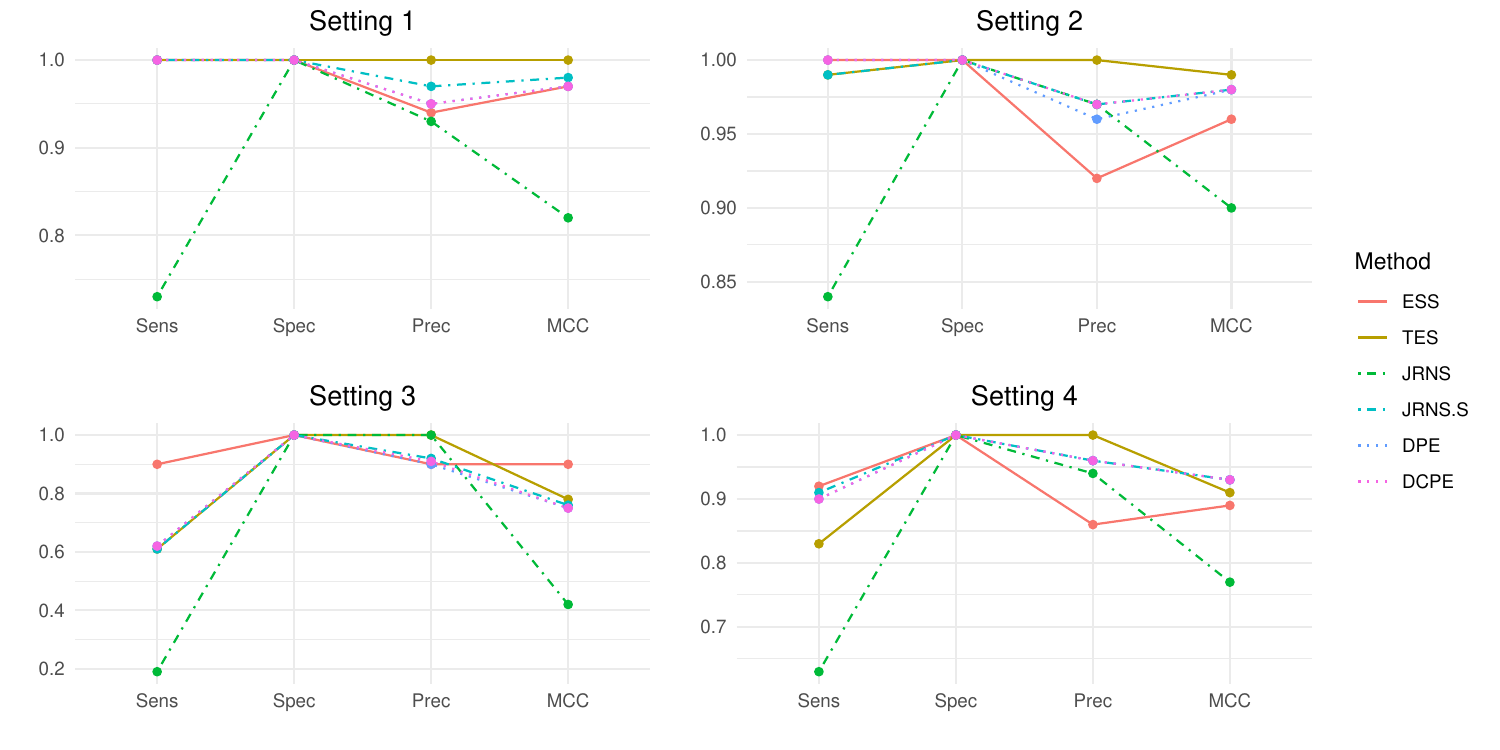}
			\caption{Sensitivity (Sens), Specificity (Spec), Precision (Prec) and MCC for sparsity selection in $B$ averaged over 10 replicates for Scenario 1 are represented for different settings of $B_0$.}
			\label{fig:table1}
		\end{figure}
%
		
		\begin{figure}[!tb]
			\centering
			\includegraphics[width=.9\linewidth]{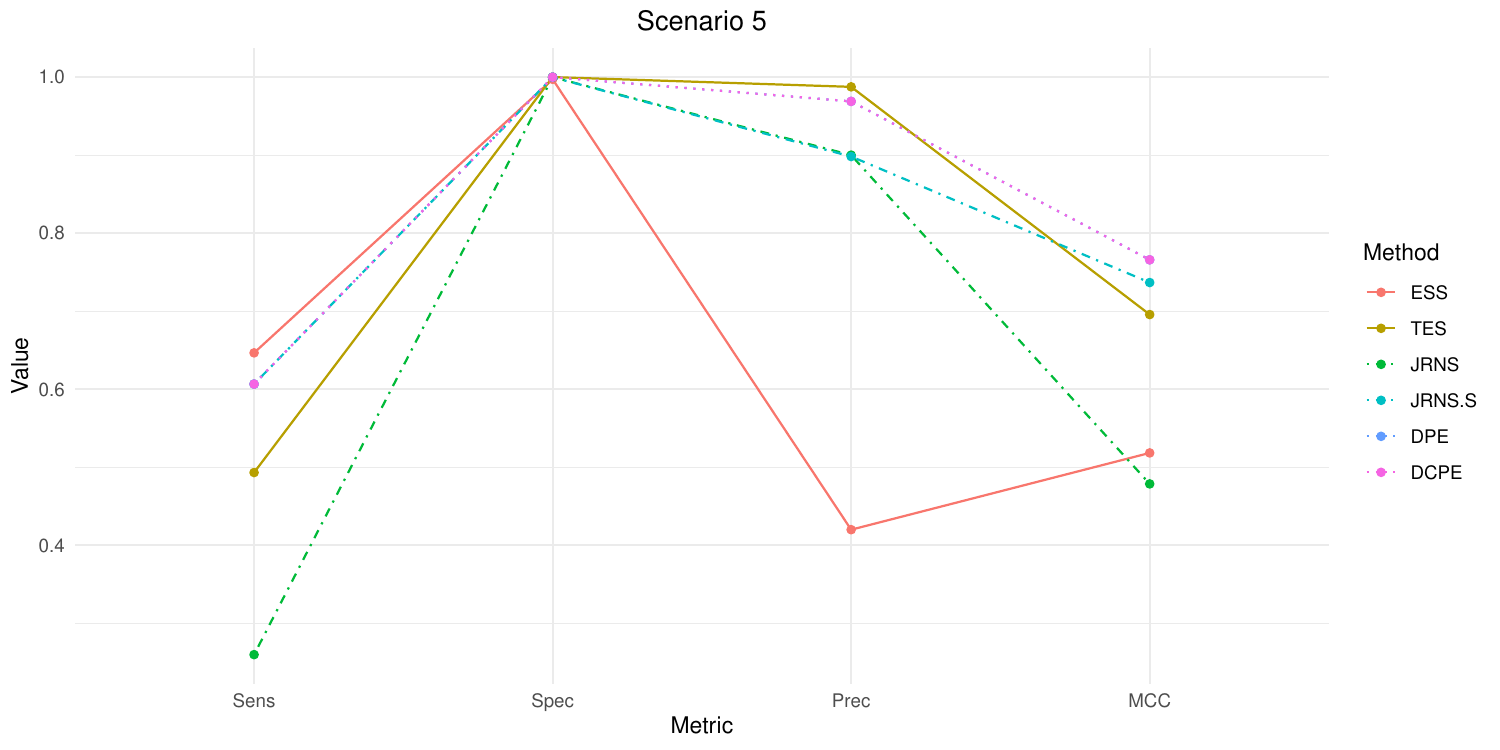}
			\caption{Sensitivity (Sens), Specificity (Spec), Precision (Prec) and MCC for sparsity selection in $B$ averaged over 10 replicates for Scenario 5 are represented.}
			\label{fig:tableS5}
		\end{figure}

		 Figures \ref{fig:table1} and \ref{fig:tableS5} represent the variable selection performance for $B_0$ of the methods in Scenarios 1 and 5, respectively.
		The results for other scenarios are provided in the supplementary material, as they show similar trends.  
		Based on the results, the performance of both ESS and TES is competitive with the other methods. 
		Specifically, TES outperforms other methods in Settings 1 and 2 across all the scenarios.
		This is also in line with our technical assumption (A2) that sets an lower bound on the minimum signal size of the true regression coefficients. 
		\begin{figure}[!tb]
			\centering
			\includegraphics[width=.9\linewidth]{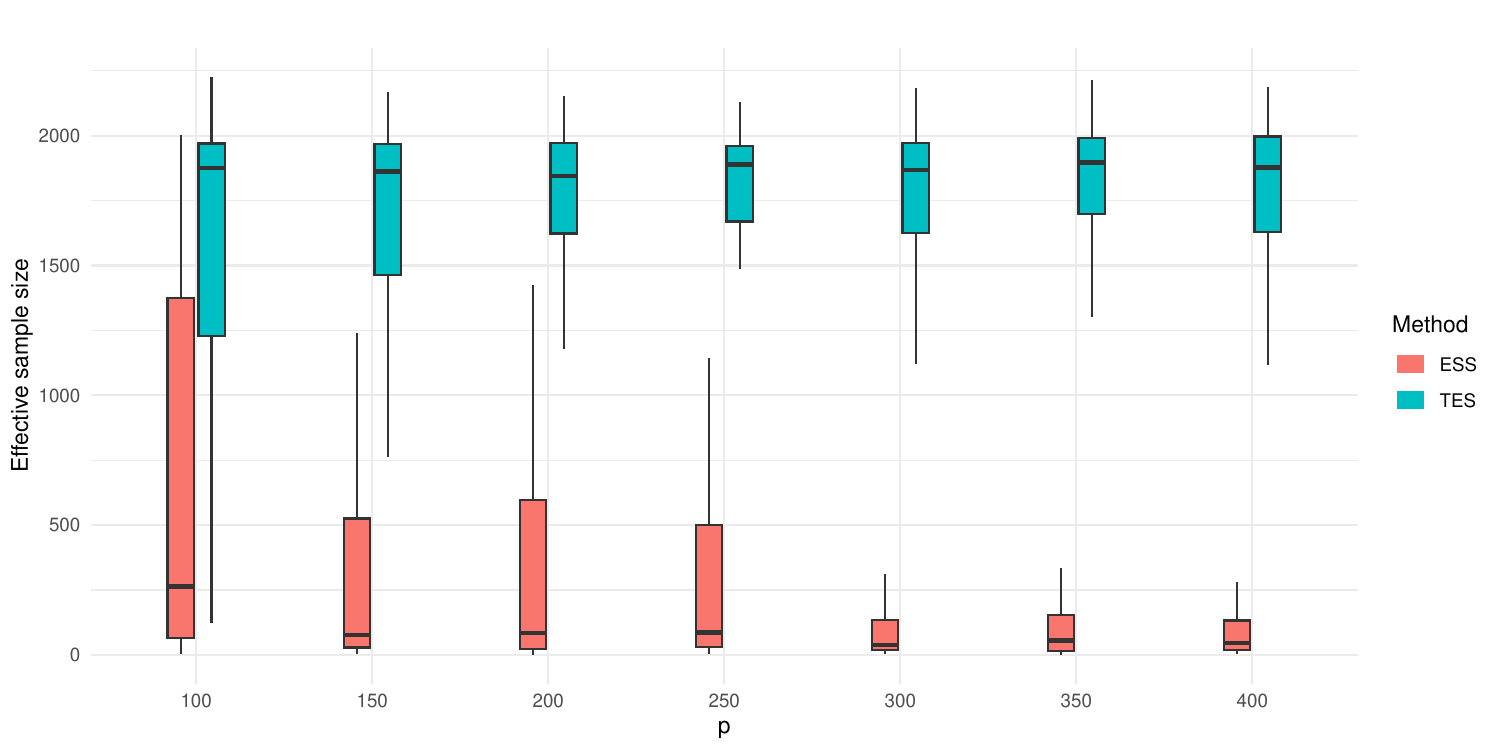}
			\caption{Comparison of the effective sample size between ESS and TES for the nonzero entries of $B_0$, $L_0$, and $D_0$ in Setting 1 and Scenario 1.}
			\label{fig:ESS}
		\end{figure}		
		Contrary to the expectation that ESS would outperform TES due to using the exact likelihood, this was not the case in our simulation studies.
		Note that ESS samples both $B$ and $\Omega$ simultaneously, while TES splits the sampling into two separate chains.
		Therefore, ESS may require longer chains to converge because it estimates more parameters, which could explain the better performance of TES in Settings 1 and 2.
		To support this claim, we compared the effective sample sizes \citep{vehtari2021rank} of ESS and TES under Setting 1 and Scenario 1 while varying the dimension $p$.
		Figure \ref{fig:ESS} presents boxplots of the effective sample sizes for the nonzero entries of $B_0$, $L_0$ and $D_0$ when $p \in \{100, 150, 200, 250, 300, 350, 400\}$.
		It shows that ESS has a significantly smaller effective sample size than TES, especially in high-dimensional settings, implying that more MCMC samples may be needed to achieve comparable posterior inference performance.

		\begin{figure}[!tb]
			\centering
			\includegraphics[width=.9\linewidth]{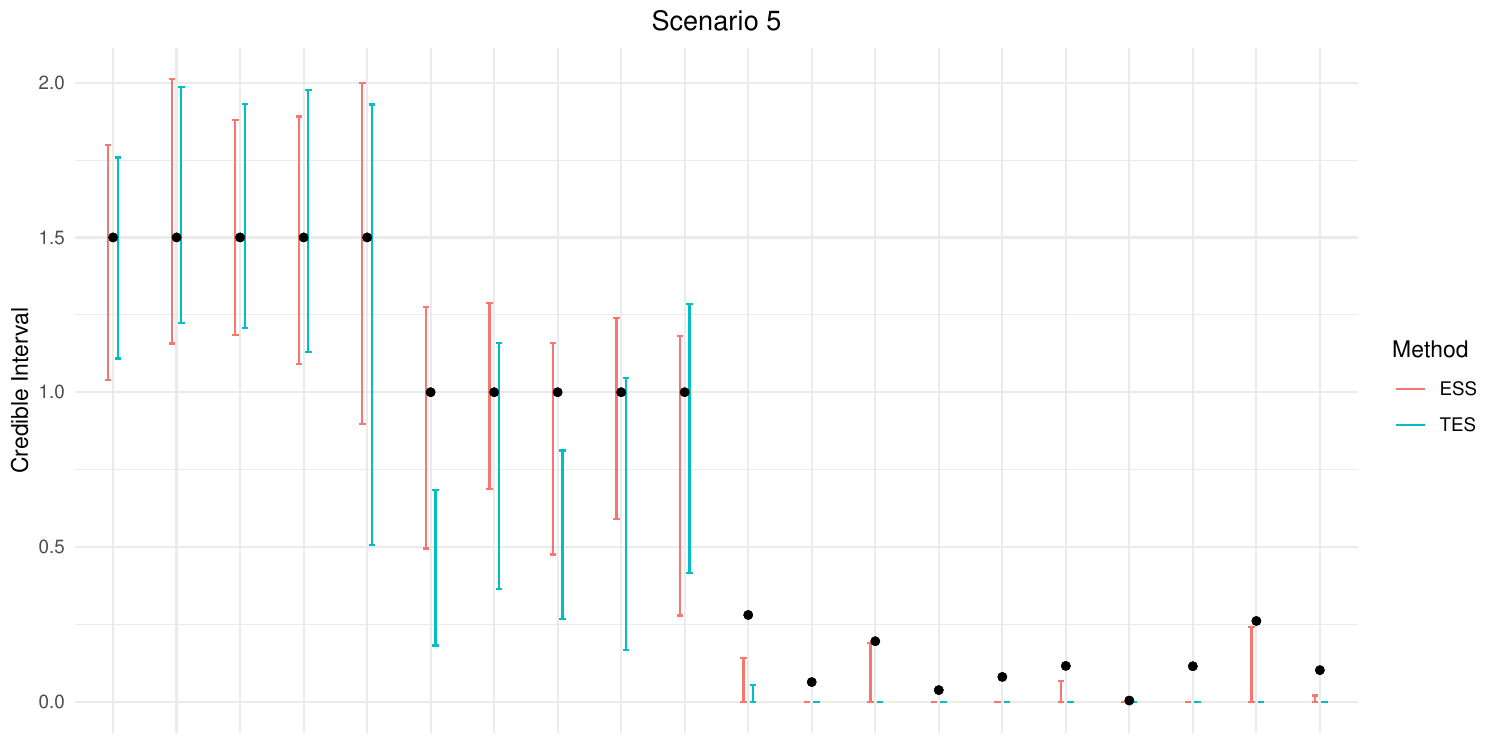}
			\caption{95\% credible intervals for the nonzero entries of $B_0$ in Scenario 5 are represented. The black dots represent the true nonzero values.}
			\label{fig:S5_CI}
		\end{figure}
		
		However, in relatively weaker signal settings (Settings 3 and 4), ESS tends to be more accurate than TES with higher sensitivity values.
		Especially in Setting 3, which contains the smallest signals among all settings, ESS achieves higher MCC values than TES, except in Scenario 3.
		This can be interpreted as the ability of joint modeling to capture the dependencies between variables, allowing for relatively better detection of small signals.
		This pattern is also observed in Figure \ref{fig:S5_CI}, where 95\% credible intervals for nonzero entries of $B_0$ in Scenario 5 are represented. 
		Separate modeling (TES) captures large signals effectively, but tends to miss small signals compared to joint modeling (ESS).
		Similar observations were also reported in \cite{Deshpande2019}.
		
		

		\begin{table}[!tb]
			\centering
			\caption{
				The summary statistics for sparsity selection in $L$ averaged over 10 replicates are represented for different scenarios under Setting 1.
			}\vspace{.15cm}
			\begin{tabular}{cccccc}
				\toprule 
				\multicolumn{2}{c}{}                                 & Sensitivity & Specificity & Precision & MCC \\ 		
				\midrule 
				\multicolumn{1}{c}{\multirow{2}{*}{Scenario 1}} & ESS &0.94             &1           &0.77           &0.85     \\
				\multicolumn{1}{c}{}                           & TES &0.76             &1           &0.98           &0.86     \\ 		
				\midrule 
				\multirow{2}{*}{Scenario 2}                     & ESS &0.80             &1           &0.75           &0.74    \\
				& TES &0.57             &1           &0.99           &0.75    \\ 		
				\midrule 
				\multirow{2}{*}{Scenario 3}                     & ESS &0.92             &1           &0.46           &0.65     \\
				& TES &0.76             &1           &0.97           &0.86     \\ 		
				\bottomrule 
			\end{tabular}
			\label{table:comp:Lselection}
		\end{table}
		We also present the results comparing the performance of (i) the selection of the Cholesky factor $L$ and (ii) the estimation of the precision matrix $\Omega$.
		The comparison of selection performance for $L$ between ESS and TES under Setting 1 of $B_0$ for different scenarios is presented in Table \ref{table:comp:Lselection}.
		In terms of sparsity selection in $L$, the performance of both ESS and TES is comparable.
		ESS renders higher values of sensitivity compared with TES, which could be caused by the two-step approach ignoring the dependence structure among the error terms. 
		However, overall, TES shows higher precision and MCC values compared with ESS by producing less false positives.

		\begin{figure}[!tb]
			\centering
			\includegraphics[width=.9\linewidth]{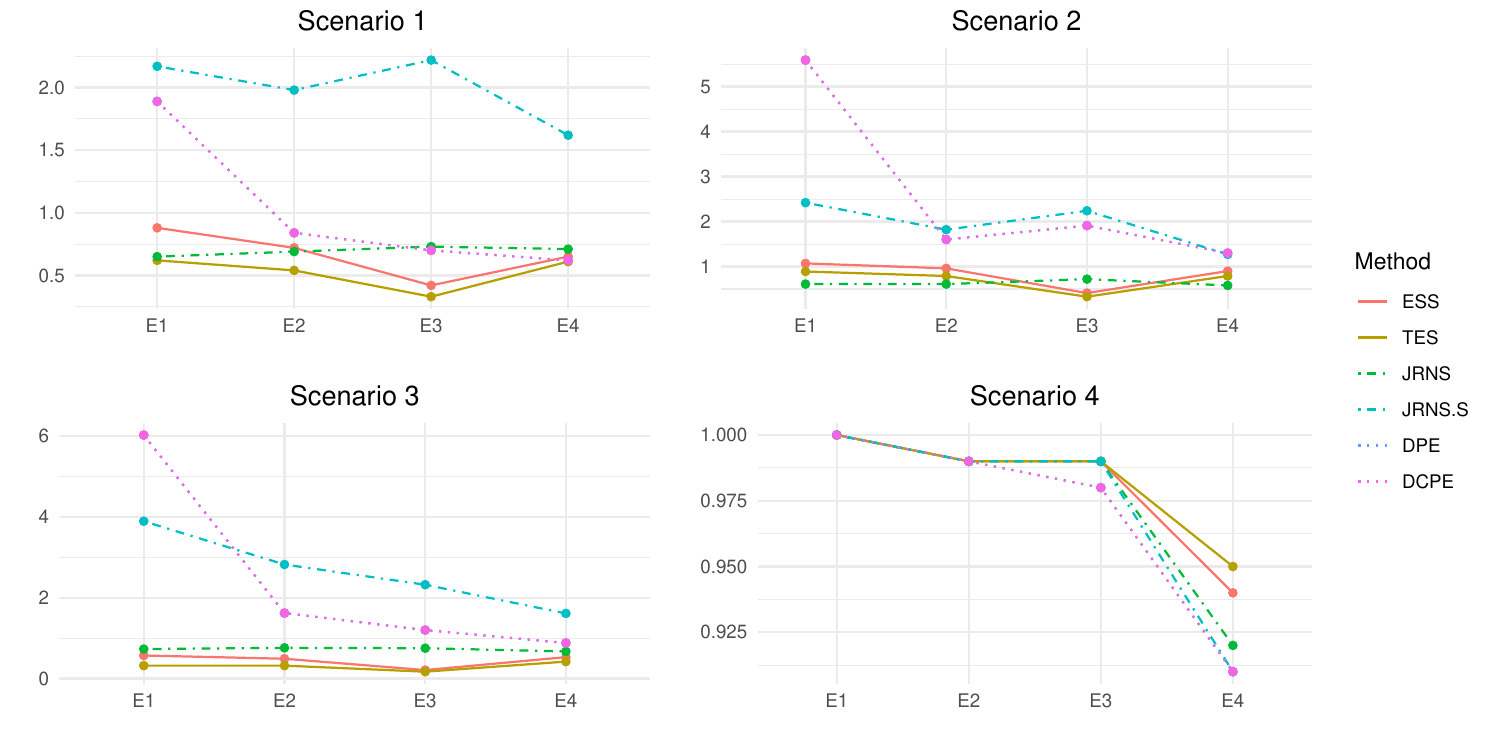}
			\caption{Relative errors in the estimation of $\Omega$ averaged over 10 replicates for Setting 1 are represented for different scenarios.}
			\label{table:comp:Oestimation}
		\end{figure}
		
		The summary statistics for estimating $\Omega$ is given in Figure \ref{table:comp:Oestimation} under the Setting 1 of $B_0$ for different scenarios. 
		In the figure, $E_1$, $E_2$, $E_3$ and $E_4$ represent the relative errors based on the matrix $\ell_1$-norm, the matrix $\ell_2$-norm (spectral norm), the vector $\ell_2$-norm (Frobenius norm) and the vector $\ell_\infty$-norm (entrywise maximum norm), respectively.
		We define the relative errors as $\|\hat{\Omega}-\Omega_0\| / \|\Omega_0\|$, where $\|\cdot\|$ denotes a norm. 		
		When estimating $\Omega$, ESS and TES perform comparably well in all scenarios including both directed and undirected graphs. 
		Especially when the underlying graph is a DAG (Scenarios 1--3), ESS and TES outperform other methods.

		In the supplementary material, we have presented additional simulation studies investigating (i) the selection and estimation performance for $B$, 
		(ii) the impact of misspecified ordering on graph estimation, 
		(iii) the uncertainty quantification performance of the proposed methods, 
		(iv) a runtime comparison of the various methods and 
		(v)  the impact of the empirical sparse prior on the performance of TES.

		\section{Real Data Analysis: assessing temperature factors in the northwest region} \label{sec:real}

		\begin{figure}[!tb]
			\centering
			\includegraphics[width=9cm]{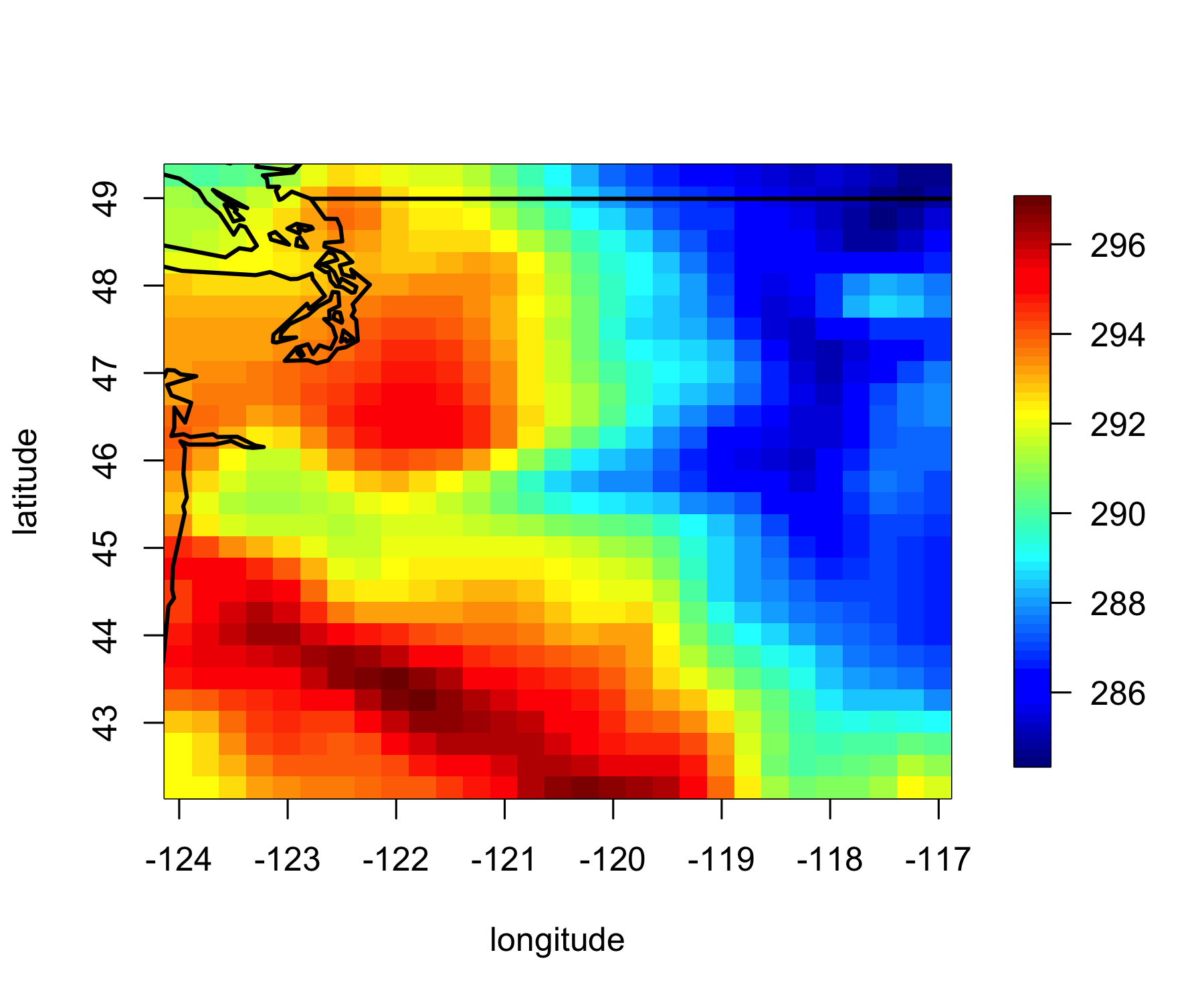}
			\caption{The temperature heatmap overlaid with the world map.
			}\label{fig:temp_map}
		\end{figure}
		
		The European Centre for Medium-Range Weather Forecasts (ECMWF) provides the ERA5 dataset, which represents the fifth generation of ECMWF's atmospheric reanalysis of the global climate, covering the period from January 1940 to the present. ERA5 offers hourly estimates of a wide range of climate variables on a 30km grid, with atmospheric resolution extending across 137 levels from the surface up to an altitude of 80km. 
		  For our analysis, we extracted temperature data for October 2023 (31 days, $q=31$) at the 1000 hPa pressure level. 
		The data were collected from 100 randomly chosen locations ($n=100$) primarily in the states of Washington and Oregon, spanning latitudes from $42.26^\circ$N to $49.26^\circ$N and longitudes from $124.01^\circ$W to $117.01^\circ$W (Figure \ref{fig:temp_map}). 
		For each location, we extracted the monthly averages of humidity, rainwater content, and wind speed for September 2023.
		In this case study, the responses are the temperature data for the 31 days, with temporal dependency providing a natural ordering of the responses. 
		The predictors include latitude, longitude, humidity, rainwater content, wind speed, and elevation. To evaluate the accuracy of different methods, we also simulated 100 noisy predictors and added them to the data alongside these six predictors ($p=106$).

		We applied a multivariate linear regression model to the data, and the proposed method and its competitors were used to estimate the coefficient matrix. 
		Note that even after adjusting for location-specific predictors, residual temporal dependencies between days may remain due to slowly evolving atmospheric conditions, such as persistent cloud cover or regional weather systems that span multiple days. 
		The proposed methods capture these temporal dependencies by imposing a Gaussian DAG model on the error vector $E_i$. 
		Since the 31 temperature variables correspond to consecutive calendar days, the assumption of a known parent ordering is both reasonable and scientifically well-motivated in the context of this empirical study.

		
			\begin{figure}[!tb]\centering
			\centering
			\includegraphics[width=1.0\linewidth]{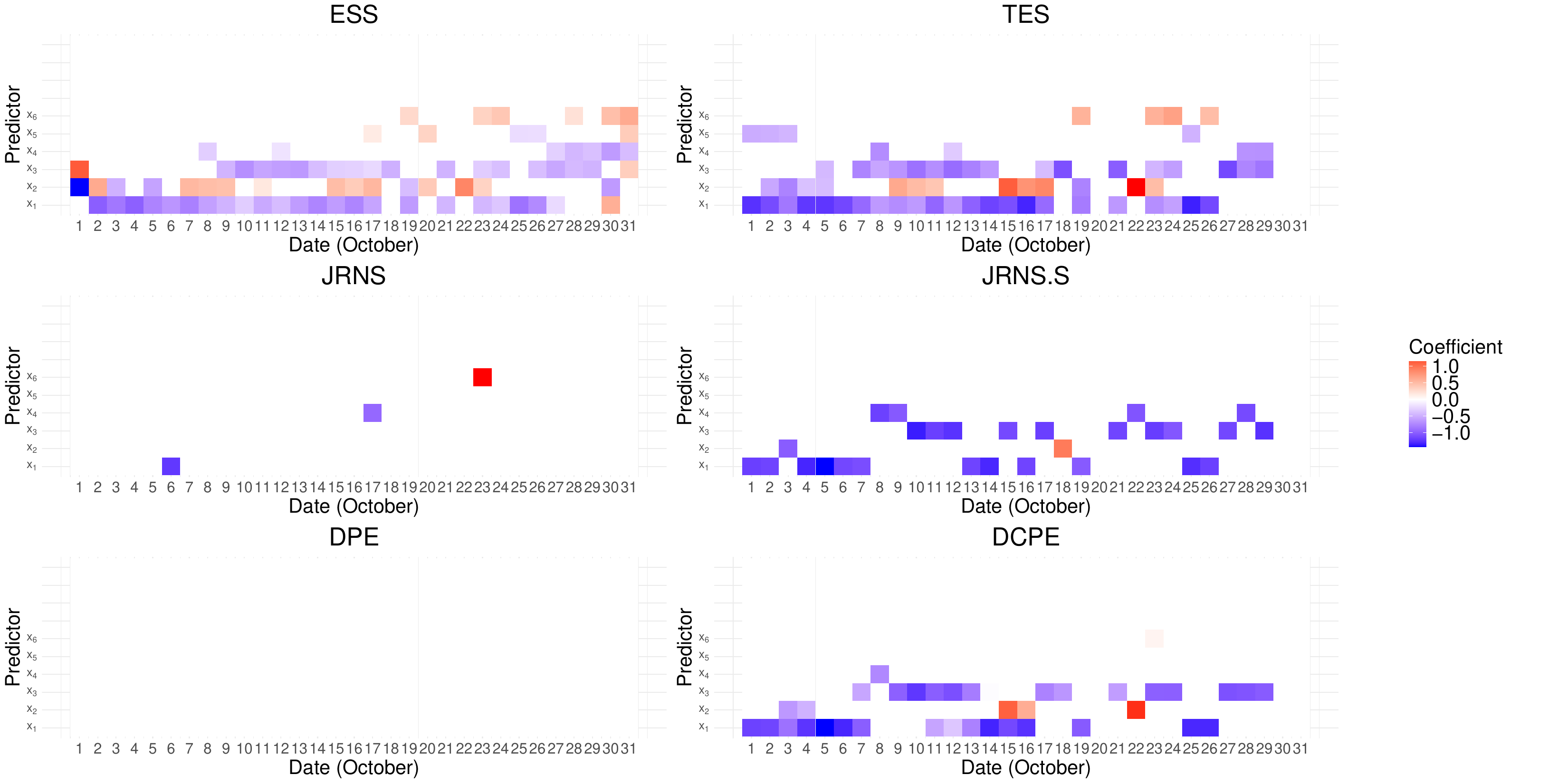}
			\caption{
				Comparison of the estimated coefficient matrices by different methods.
			}\label{fig:est_B_real}
		\end{figure}
		The regression coefficient matrices estimated by each method are shown in Figure \ref{fig:est_B_real}. 
		All methods successfully exclude all noisy predictors; therefore, only the first ten rows of the estimated coefficient matrices are presented. 
		Compared to contenders, both ESS and TES successfully identify all relevant predictors with high precision and sensitivity. 
		In particular, both ESS and TES identify latitude (which is the covariate at the bottom of Figure \ref{fig:est_B_real}) as being negatively associated with temperature, reflecting the well-known geographical pattern where temperatures decrease from southwest to northeast in the Pacific Northwest. 
		This decrease in temperature with increasing latitude is primarily due to factors such as the angle of solar radiation, proximity to polar regions, and the influence of ocean currents and atmospheric circulation patterns. 
		This trend is clearly shown in Figure \ref{fig:temp_map}, where decreasing temperatures correlate with increasing latitude. 
		Additionally, both humidity and rainwater content (the third and fourth covariates from the bottom in Figure \ref{fig:est_B_real}, respectively) are estimated to be negatively associated with temperature. 
		The relationship between temperature and rainfall in Washington during October is influenced by the diverse geography of the state and prevailing weather patterns. As temperature decreases, the capacity of the atmosphere to hold moisture diminishes, often leading to increased precipitation.
		This transition is particularly evident in Western Washington, where the onset of the rainy season corresponds with cooler temperatures. 
		Wind speed (the fifth covariate from the bottom in Figure \ref{fig:est_B_real}) is also shown to negatively affect temperature, particularly in the winter time. This effect is due to enhanced mixing of air layers, which brings in colder air, and the promotion of cold air advection. 
		These observations clearly demonstrate the advantages of the proposed methods for precise parameter estimation when applied to real datasets.

		\begin{figure}[!tb]\centering
			\centering
			\includegraphics[width=\linewidth]{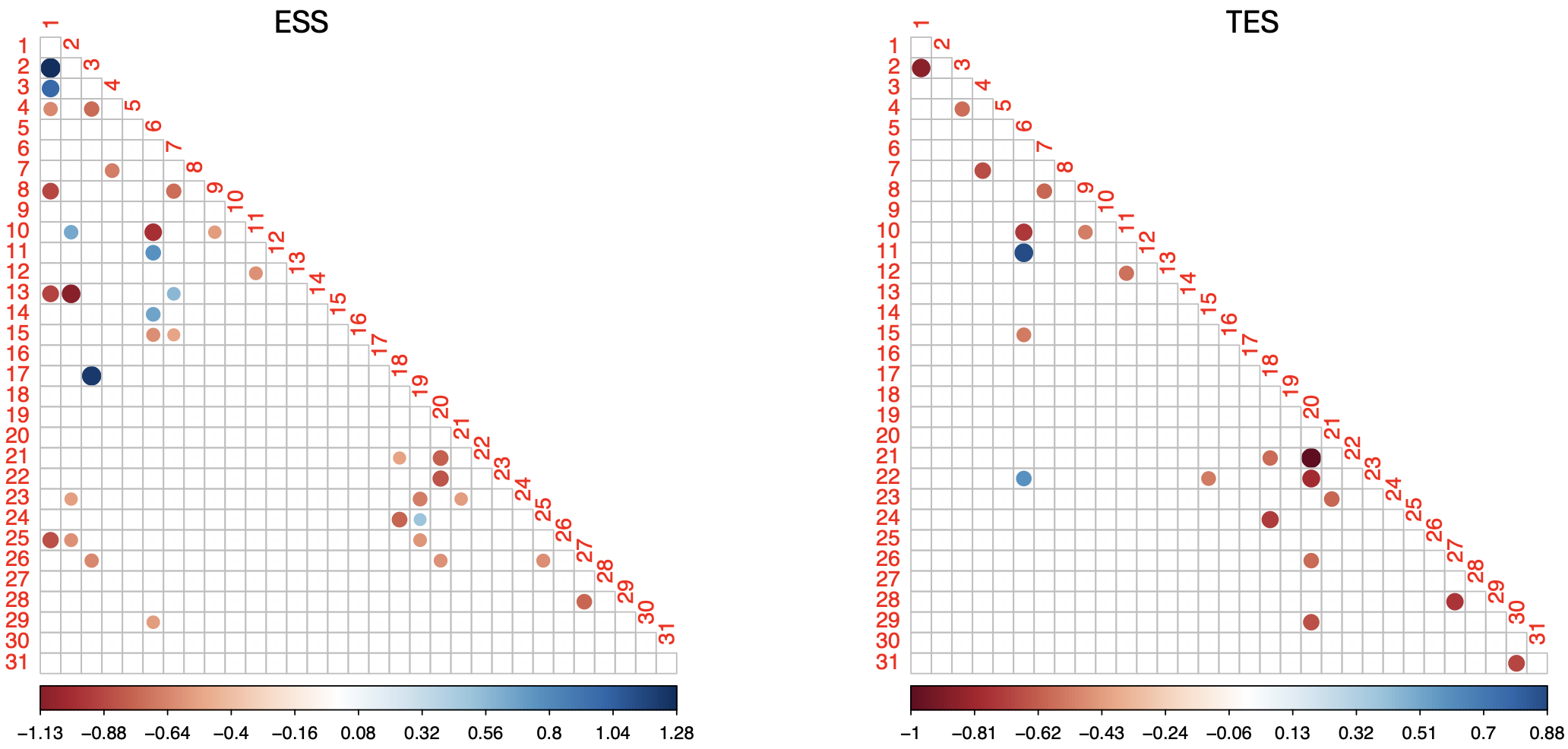}
			\caption{
				Comparison of the estimated Cholesky factors based on ESS (left) and TES (right) using an absolute cutoff of 0.5.
			}\label{fig:est_L_real}
		\end{figure}
		To assess how well ESS and TES capture residual temporal dependencies, we also investigated the estimated Cholesky factors $L$.
		Figure \ref{fig:est_L_real} shows only the off-diagonal entries of the estimated Cholesky factors whose absolute values exceed 0.5.
		As shown in the figure, most of the significant entries are negative, corresponding to positive coefficients in the regression formulation of DAG models. 
		In addition, these significant entries tend to be located near the diagonal, which is consistent with the expectation that higher temperatures on nearby previous days have a positive effect on the temperature of the current day.


		\section{Discussion} \label{sec:disc}
		
		In this paper, we consider the joint estimation of the regression coefficient matrix and the error precision matrix in high-dimensional multivariate linear regression models. 
		The two Bayesian methods, the exact likelihood-based (ESS) and two-step (TES) approaches, are proposed.
		Here, we offer the following guidelines for the two proposed methods to help readers choose the appropriate method for their specific situation.		
		If inference on small signals is crucial or the dimensions ($p$ and $q$) are relatively small, we recommend using ESS. In other cases, we generally recommend TES as the preferred method.
		As shown in our simulation results, TES demonstrates reasonable variable selection performance in many settings, excluding small signals, and is significantly faster than ESS. 
		Even in cases with small signals, TES offers a meaningful speed improvement with a slight sacrifice in performance.

		There are several possible directions for future work. 
		First, in the proposed two-step method, the penalty for model sparsity is controlled by the hyperparameter $c_1$. 
		To adaptively select an appropriate $c_1$ value based on the true sparsity of the coefficient matrix $B$, adding another layer of prior on $c_1$ could potentially make the model more robust and less sensitive to the tuning parameter.
		Second, it would be interesting to investigate whether one can obtain selection consistency for ESS corresponding to the exact likelihood. 
		This would need a significant amount of technical modification, so we leave it as future work. 
		Finally, the assumption of a known ordering under a directed graph can be restrictive in certain applications. 
		While the parent ordering assumption enables scalable inference by avoiding the combinatorial complexity of searching over all possible DAGs, it limits the proposed method to explore only those DAGs that are compatible with the given ordering. 
		An important and interesting direction for future research is to develop methods for estimating the dependence structure among responses when the ordering of variables in a DAG is unknown, as explored, for example, in \cite{consonni2017objective} and \cite{castelletti2022bcdag}.

		\begin{acks}[Acknowledgments]
			The authors would like to thank the anonymous referees, an Associate
			Editor and the Editor for their constructive comments that improved the
			quality of this paper.
		\end{acks}
		\begin{funding}
			Xuan Cao was supported by the Summer Research Fellowship from The Taft Research Center at the University of Cincinnati.
			Kyoungjae Lee was supported by the National Research Foundation of Korea (NRF) grant funded by the Korea government (MSIT) (RS-2023-00211974 and RS-2025-02216235).
		\end{funding}
		\begin{supplement}
			\stitle{Supplementary to ``Scalable Bayesian inference on high-dimensional multivariate linear regression''}
			\sdescription{It contains the proofs of the main results in this paper and additional simulation results.}
		\end{supplement}
		
		\bibliographystyle{ba}
		\bibliography{references}

\end{document}